\documentclass[12pt,preprint]{emulateapj}
\usepackage{graphicx}

\slugcomment{\today}

\shorttitle{Supernova Neutrino}
\shortauthors{Nakazato et al.}

\begin{document}


\title{Supernova Neutrino Light Curves and Spectra for Various Progenitor Stars: \\ From Core Collapse to Proto-neutron Star Cooling}


\author{Ken'ichiro Nakazato\altaffilmark{1}, Kohsuke Sumiyoshi\altaffilmark{2}, Hideyuki Suzuki\altaffilmark{1}, Tomonori Totani\altaffilmark{3}, \\ Hideyuki Umeda\altaffilmark{4} and Shoichi Yamada\altaffilmark{5,6}}

\email{nakazato@rs.tus.ac.jp}


\altaffiltext{1}{Department of Physics, Faculty of Science \& Technology, Tokyo University of Science, 2641 Yamazaki, Noda, Chiba 278-8510, Japan}
\altaffiltext{2}{Numazu Collage of Technology, 3600 Ooka, Numazu, Shizuoka 410-8501, Japan}
\altaffiltext{3}{Department of Astronomy, Kyoto University, Kita-shirakawa Oiwake-cho, Sakyo, Kyoto 606-8502, Japan}
\altaffiltext{4}{Department of Astronomy, University of Tokyo, 7-3-1 Hongo, Bunkyo, Tokyo 113-0033, Japan}
\altaffiltext{5}{Department of Physics, Waseda University, 3-4-1 Okubo, Shinjuku, Tokyo 169-8555, Japan}
\altaffiltext{6}{Advanced Research Institute for Science \& Engineering, Waseda University, 3-4-1 Okubo, Shinjuku, Tokyo 169-8555, Japan}


\begin{abstract}
We present
a new series of supernova neutrino light curves and spectra calculated
by numerical simulations for a variety of progenitor stellar masses
(13-$50M_\odot$) and metallicities ($Z = 0.02$ and 0.004), which would be
useful for a broad range of supernova neutrino studies, e.g.,
simulations of future neutrino burst detection by underground detectors,
or theoretical predictions for the relic supernova neutrino background.
To follow the evolution from the onset of collapse to 20~s after the core bounce,
we combine the results of neutrino-radiation hydrodynamic simulations for the early phase and quasi-static evolutionary calculations of neutrino diffusion for the late
phase, with different values of shock revival time as a parameter
that should depend on the still unknown explosion mechanism.
We here describe the calculation methods
and basic results including the dependence on progenitor models
and the shock revival time. The neutrino data are publicly available electronically.
\end{abstract}



\keywords{supernovae: general --- neutrinos --- stars: neutron --- black hole physics --- hydrodynamics --- methods: numerical}


\section{Introduction} \label{intro}

Supernova explosion is one of the most spectacular events in the
universe. It is not only the death of massive stars but also an engine
for the evolution of the galaxies. Apart from the classification
according to the spectroscopy, it is thought that there are two
mechanisms of the explosion. While type Ia supernova is caused by the
thermonuclear explosion of white dwarfs, the other types of supernovae
are driven by the core collapse of massive stars. Unfortunately,
however, details of the supernova explosion remain unsolved.

As for the collapse-driven supernova, the physics which makes the
explosion is not well understood. While many numerical simulations for
the gravitational collapse of massive stars have been done so far
\citep[e.g.][for recent reviews]{ott09,thiel11,kotake12,janka12}, there is no
consensus on the explosion mechanism of collapse-driven
supernovae. However, a rough sketch of the scenario shown below is
widely accepted. The core of massive star becomes gravitationally
unstable at the end of evolution and starts to collapse. The collapse
is bounced by the nuclear repulsion force and the shock wave is launched.
A supernova explosion is observed when the shock wave successfully
propagates up to the stellar surface and expels the envelope material.
Finally the neutron star or black hole is formed
as a remnant.  However, the general difficulty in numerical
simulations for successful explosions is that shock waves tend to
stall before blowing out the stellar envelopes. A possible mechanism
to revive the stalled shock waves is energy input by strong neutrino
radiation from the newborn neutron stars \citep[``the delayed explosion
scenario''; e.g.][]{bethe85}, but it is still a matter of debate whether this single
process is sufficient for successful explosions. 
An aspherical hydrodynamic turbulence, such as the convective instability and standing accretion shock instability, may help the shock revival \citep[e.g.][]{herant94,blond03}.
Other proposed
mechanism or physical processes to account for explosions include QCD phase transition \citep[][]{fischer11}, acoustic wave \citep[][]{burr06} and magnetic field \citep[][]{lebla70}.

The collapse-driven supernova is important also as a target of
neutrino astronomy. As is well known, supernova neutrinos from SN1987A
have been detected by the two water Cerenkov detectors, Kamiokande II
\citep[][]{hirata87} and IMB \citep[][]{bionta87}. Since the Large
Magellanic Cloud, where SN1987A appeared, is $\sim$50~kpc away from
the Earth and detectors at the time were not as large as the
present-day experiments, the event number was small (eleven for
Kamiokande II and eight for IMB). Nevertheless, many theoretical
studies tried to extract various information about physics of
supernovae and/or neutrinos from these data \citep[e.g.][for a recent review]{raffelt12}.  If a supernova occurs now near the Galactic center
($\sim$10~kpc away from the Earth), about 10,000 events will be
detected by SuperKamiokande, which is the largest, currently operating
neutrino detector around MeV \citep[][]{burr92,totani98}.
Therefore details of a collapse-driven supernova
such as the explosion mechanism, as well as the neutrino physics such
as mass and its hierarchy, mixing and oscillation, or any exotic
physics, could be investigated by high-statistics light curve and
spectrum of a future supernova neutrino burst event.

Another potential opportunity to observe neutrinos
from supernovae is the relic background radiation, which
gives valuable clues about e.g., the cosmic star formation
history and/or stellar initial mass function
\citep[e.g.][]{totani96,ando04,beacom10}.
The cosmic metallicity evolves with time and, as stated in Section~\ref{proge}, the stellar evolution depends not only on its mass but also on its metallicity. Since the metal-free and very metal-poor massive stars are formed in the high redshift universe and explode immediately, their contribution to the relic neutrino background would be small \citep[][]{ando04}. However, galaxies whose metallicities are about one order of magnitude lower than the solar value reside within the redshift $z \lesssim 1$ \citep[][]{peesom12}.
To predict the flux and spectrum of the relic supernova
neutrinos, one must model the integrated spectrum of
supernova neutrinos for various ranges of progenitor stars.

Considering these situations, it would obviously be useful to provide the
results of supernova neutrino emission calculated by state-of-art
numerical simulations, for various types of progenitor stars. Such a
comprehensive datab
ase of theoretical predictions can be used as
templates to simulate supernova neutrino detection of a future
neutrino burst event, or to construct a realistic prediction for the
relic neutrino background. However, most of numerical simulations so
far followed only within $\lesssim$1~s \citep[e.g.][]{marek09} after
the core collapse because solving the neutrino transfer with
hydrodynamics for a long term is a numerically tough problem.  Such
simulations are not satisfactory for the above purpose, because a
significant fraction of the total gravitational energy is emitted as
neutrinos after this early phase, with a typical decay time scale of
$\sim$10~s. Such a simulation result for a long term was shown in
\citet{totani98}. Recently, \citet{fischer10,fischer12} showed
long term neutrino signals for two models, an ordinary collapse-driven
supernova and an electron-capture supernova, which is a subcategory of
collapse-driven supernovae. As for the electron-capture supernova,
long term simulations were also performed by \citet{hude10} computing 
neutrino signals. On the other hand, neutrino diffusions in
the late phase were separately investigated by evolutionary
calculations \citep[e.g.][]{burr86,sumi95,pons99,robe12}.  However, there is no
comprehensive data set yet for numerical long-term supernova neutrino
signals from various types of progenitor stars.

In this study, we construct a publicly available database for
numerical simulations of supernova neutrino emission, including eight
simulations for normal single stars in the initial mass range of
13-$50M_\odot$ for two different values of stellar metallicity ($Z=0.02$
and 0.004, i.e., the solar abundance and its $1/5$, respectively). To
follow the long-term evolution for many simulations, we use two
different simulation methods: general relativistic neutrino-radiation
hydrodynamic ($\nu$RHD) simulations for the early phase, and general
relativistic quasi-static evolutionary calculations of neutrino diffusion in a
nascent neutron star for the late phase.  Though the $\nu$RHD
simulations do not lead to a natural supernova explosion, we introduce
a parameter, shock revival time, which reflects the unknown explosion
mechanism, and phenomenologically connect the early and late phases by
this parameter based on physical considerations.  Although the
connection between the two phases is not perfectly consistent as a
physical simulation, this approach allows us to provide reasonably
realistic supernova neutrino light curves and spectra from the onset
of a core collapse to $\sim$10~s after that, for various progenitor
stellar models.

The organization of this paper is as follows. In Section~\ref{proge},
the progenitor models used in this study are described.  The early
phase of supernovae, from the onset of collapse to the shock
propagation, is described in Section~\ref{colla}.  In
Section~\ref{pnsc}, the quasi-static evolutionary calculations of
neutrino diffusion in the dense core for the late phase are described.
The connection between the early and late phases are discussed in
Section~\ref{neusg}.  Section~\ref{disc} is devoted to a summary and
discussion.

\section{Progenitors of collapse-driven supernova} \label{proge}

Massive stars with the initial mass $M_\mathrm{init}\gtrsim 10M_\odot$ are known to undergo gravitational collapse at the end of their lives. They synthesize heavy elements through some nuclear burning stages during the quasi-static evolutions. Finally the iron core, where all nuclear reactions become equilibrium to inverse reactions of themselves and various elements called ``iron'' (Fe, Ni, Co, Mn and so on) are created, is formed at the center. The iron core has density of $\gtrsim$10$^{10}$~g/cm$^3$ and is supported by the degenerate pressure of relativistic electrons. On the other hand, the condition for the gravitational instability is written, using the adiabatic index, $\gamma$, as
\begin{equation}
\gamma \equiv \left(\frac{\partial \ln p}{\partial \ln \rho}\right)_s < \frac{4}{3},
\label{adia}
\end{equation}
where $p$, $\rho$ and $s$ are the pressure, density and entropy, respectively. Since the adiabatic index of relativistic ideal gas is $4/3$, the iron core is marginally stable.

There are two processes which destabilize the iron core. The first one is an electron capture by protons belonging to nuclei:
\begin{equation}
p + e \longrightarrow n + \nu_e.
\label{ecp}
\end{equation}
This reaction is caused because the sum of rest mass and kinetic energy of the relativistic electrons exceeds the mass difference between neutron and proton due to high density. Diminution of electrons causes a deficit in pressure and makes the iron core unstable. The second is the photodisintegration reaction of nucleus such as
\begin{equation}
^{56}\mathrm{Fe} \longrightarrow 13 \,^4\mathrm{He} + 4 \, n \longrightarrow 26 \, p + 30 \, n.
\label{photodis}
\end{equation}
This process works for high temperature as $T \sim 5 \times 10^9$~K. While the energy per nucleon, $\varepsilon$, reaches a minimum at $^{56}$Fe, the Helmholtz free energy per nucleon, ${\cal F}$, is minimized in finite temperature. Because of ${\cal F}=\varepsilon-Ts$, matter becomes thermodynamically stable by the increase in entropy due to the photodisintegration of nuclei. Nevertheless, this is an endothermic reaction so that the pressure does not increase so much as to suspend the core collapse. This is an onset of the collapse-driven supernova.

\begin{figure*}
\plotone{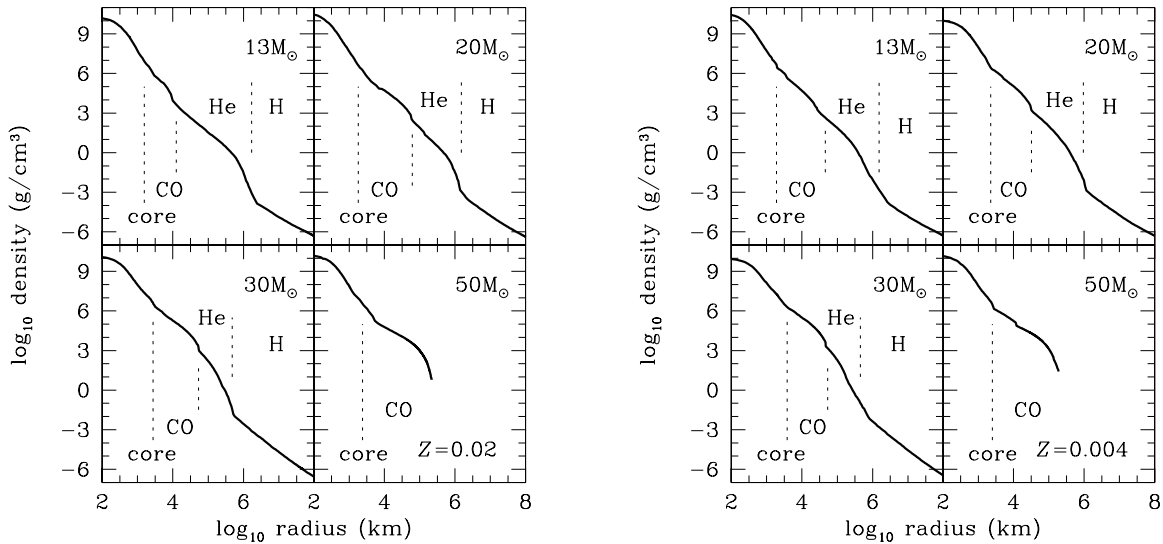}
\caption{Density profiles of progenitor models with the  metallicity $Z=0.02$ (left panel) and $Z=0.004$ (right panel). In both panels, upper left, upper right, lower left and lower right plots corresponds to the models with the initial mass $M_\mathrm{init}=13M_\odot$, $20M_\odot$, $30M_\odot$ and $50M_\odot$, respectively. Dotted vertical lines are shown to guide eyes.}
\label{inidnsall}
\end{figure*}

The lower mass limit for the progenitors of collapse-driven supernovae is determined whether the iron core is formed or not. Note that, for somewhat low mass cases, the core collapse may be triggered only by electron captures before Ne ignition, which is called an electron-capture supernova. They can be regarded as a subcategory of collapse-driven supernovae because their scenarios after the collapse are thought to be similar. \citet{poe08} investigated stellar evolution sequences with initial masses between 6.5 and $13M_\odot$ and the solar metallicity using three different codes and found that the lower mass limit for core collapse is 9-$12M_\odot$. Observationally, \citet{smartt09} found that it converges to $8 \pm 1M_\odot$ investigating the collapse driven supernovae observed in a fixed 10.5-year period within a distance of 28~Mpc.

The upper mass limit is determined whether the supernova explosion succeeds or not. It is theoretically uncertain because the precise understanding on the explosion mechanism is still absent. Nevertheless, generally to say, the density profile of progenitor is important because the shock wave should run through the star. What makes a situation more awkward is uncertainty in the theory of stellar evolution, the mass loss rate and convection. The effects of rotation and binary interaction may also affect the stellar evolution. \citet{fryer99} showed numerically that nonrotating stars with initial masses $\gtrsim \!\!40M_\odot$ fail to explode and form black holes directly while the mass loss was not considered in the progenitor models adopted by them. Note that, the black holes could be also formed after the explosion, which is probably weak, via fallback accretion. Recently, \citet{ocon10} performed the core collapse simulations for many progenitor sets with the spherically symmetric (1D) models involving simplified neutrino transfer. They implied that the upper mass limit for the explosion depends severely on the evolutionary calculations, while their estimation was based on the calculations with an artificially increased energy deposition.

Observational constraints on the upper mass limit are also obscure. \citet{smartt09} suggested that the maximum mass for the progenitors of type II-P supernovae is $\sim \!\!20M_\odot$ and the majority of more massive stars may collapse quietly to black holes where the explosions remain undetected. However, \citet{smith11}, who investigated the observed fractions of the collapse-driven supernovae, claimed that they will produce other types of supernovae which are not II-P. On the other hand, massive stars with the mass of $\gtrsim \!\!35M_\odot$ lose the outer layer due to their strong stellar winds and form Wolf-Rayet stars, where the helium cores are exposed \citep[][]{woosley02}. They are thought as progenitors of type Ib and Ic supernovae and, according to the light curve models by \citet{nomoto06}, the progenitors of some luminous type Ic supernovae called hypernovae have the masses of $\sim \!\!40M_\odot$. They implied that the fate of massive stars beyond $25M_\odot$ depends on the stellar rotation and the hypernova progenitors are rotating. Note that, \citet{smith11} suggested that a certain fraction of such massive stars produce type IIn supernovae from luminous blue variables. They also pointed out the importance of binary interactions to account for type Ib and Ic supernovae. Recently, extremely luminous supernovae such as SN2006gy (type IIn) and SN2007bi (type Ic) are observed and they are suggested to have a progenitor mass of $\gtrsim \!\!100M_\odot$, while they may be pair-instability supernovae, whose explosion mechanism is different from that of the collapse-driven supernovae. It should be emphasized that, as described above, the fate of massive stars is a hot subject under active discussion, but we deal progenitors provided from a single treatment for the stellar evolution in below.

\begin{figure*}
\plotone{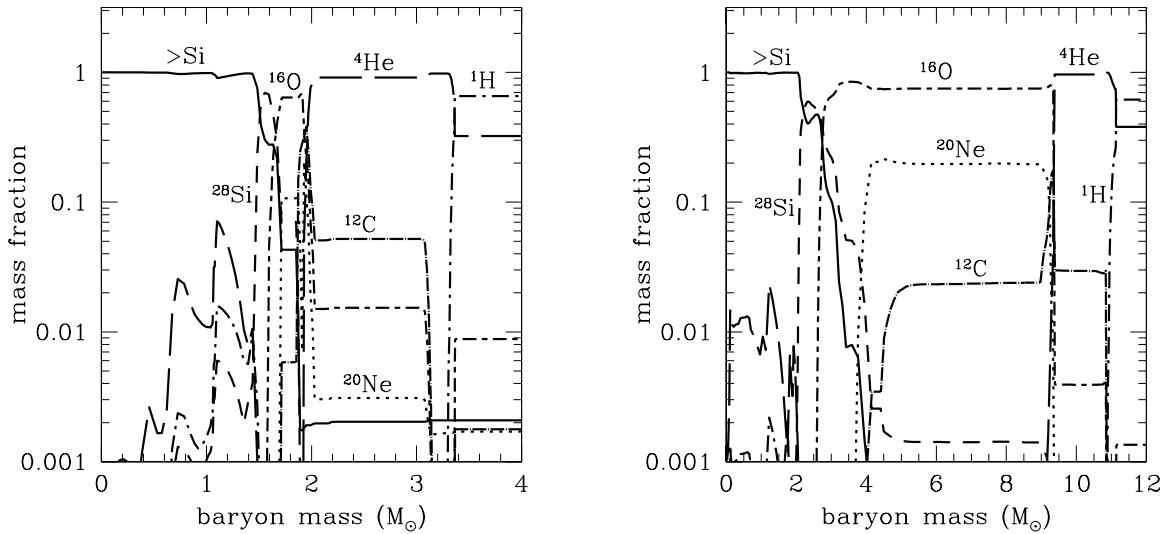}
\caption{Composition profiles of progenitor models with the initial mass $M_\mathrm{init}=13M_\odot$ and the metallicity $Z=0.02$ (left panel) and $M_\mathrm{init}=30M_\odot$ and $Z=0.004$ (right panel). Plots are shown as a function of the baryon mass coordinate (enclosed mass). In both panels, ``$>$Si'' means a sum for the elements heavier than silicon.}
\label{inicom}
\end{figure*}

In this study, we prepare eight progenitor models with the initial masses $M_\mathrm{init}= 13M_\odot$, $20M_\odot$, $30M_\odot$ and $50M_\odot$, and the metallicities of $Z=0.02$ (solar) and 0.004 (Small Magellanic Cloud). They are computed by a Henyey type stellar evolution code which is fully coupled to a nuclear reaction network. This code is also used in \citet{umeda08} and its descriptions, such as the treatment of convection, are given in \citet{umeda12}. As for the mass loss rate, we adopt the same model with ``Case A'' of \citet{umeda11} for the main sequence stage \citep[see also,][]{jager88, vink01}. We set the empirical mass loss rate \citep[][]{jager88} scaled with the metallicity as $(Z/0.02)^{0.5}$ for the Wolf-Rayet stage \citep[][]{kudrit89}. The effects of rotation and binary interaction are not taken into account.

\begin{figure}[b]
\plotone{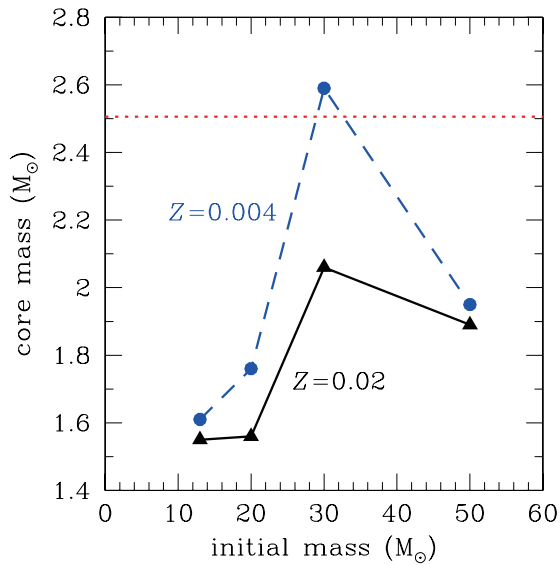}
\caption{Core mass of progenitor models as a function of the initial mass. Solid and dashed lines correspond to the cases with the metallicity $Z=0.02$ and 0.004, respectively. The horizontal dotted line represents the maximum baryonic mass of neutron stars for the equation of state by \citet{shen98a,shen98b}.}
\label{coremas}
\end{figure}

\begin{figure*}
\plotone{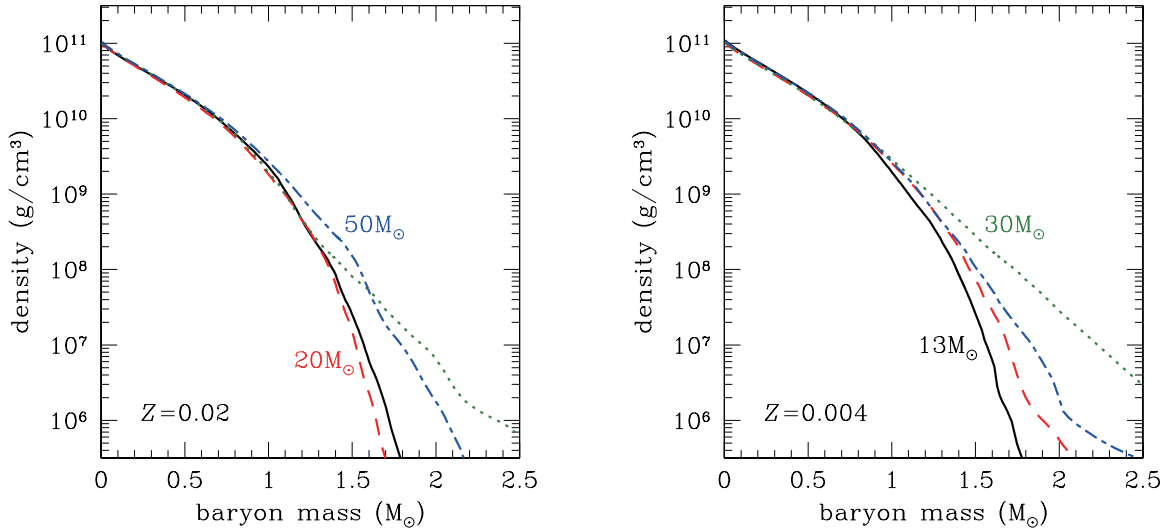}
\caption{Density profiles at the times with the central density of $10^{11}$~g~cm$^{-3}$ for progenitor models with the metallicity $Z=0.02$ (left panel) and 0.004 (right panel). In both panels, solid, dashed, dotted and dot-dashed lines correspond to the models with the initial mass $M_\mathrm{init}=13M_\odot$, $20M_\odot$, $30M_\odot$ and $50M_\odot$, respectively.}
\label{inidns}
\end{figure*}

The density profiles of our progenitor models are shown in Figure~\ref{inidnsall} and the composition profiles for two reference cases, $(M_\mathrm{init}, Z)=(13M_\odot, 0.02)$ and $(30M_\odot, 0.004)$, are shown in Figure~\ref{inicom}. Note that, as shown in Table~\ref{keyprm}, the total progenitor masses when the collapse begins, $M_\mathrm{tot}$, are different from $M_\mathrm{init}$ due to the mass loss. The progenitors have the onion-like structure: central core is surrounded by shells of lighter elements.
We regard the region of oxygen depletion as a core hereafter. In Figure~\ref{coremas}, the core mass is plotted as a function of the initial mass for all models. Since the mass loss rate is larger for high metallicity, the models with $Z=0.004$ have higher core mass than those with $Z=0.02$. The models with $M_\mathrm{init}= 50M_\odot$ become the Wolf-Rayet stars losing large amount of their mass and final core masses are somewhat low, which may correspond to type Ic supernovae. As a result, the core mass of the model with $M_\mathrm{init}= 30M_\odot$ is the highest for each metallicity case.

To evaluate the spectra of neutrinos emitted from the collapse of above progenitor models, which is the goal of this paper, we put the assumptions on the fate of our progenitors. For the equation of state by \citet{shen98a,shen98b}, which is adopted below in our simulations, the maximum mass of neutron stars is $2.2M_\odot$ in the gravitational mass and $2.5M_\odot$ in the baryonic mass. We assume that the model with $(M_\mathrm{init}, Z)=(30M_\odot, 0.004)$ forms a black hole, because its core mass is larger than the maximum baryon mass. While whether the other models can make explosions or not is unclear, we regard them as ``supernova'' progenitors. The correspondence between the initial mass and core mass is not well-established, however, the outer region far enough away from the core does not influence the dynamics and neutrino emission of the collapse. Therefore, we believe that the systematics discussed in this study will be helpful for other progenitor sets if their core masses are known.

\section{Dynamics of core collapse and bounce} \label{colla}

In this section, we follow the early stage of the collapse-driven supernova showing the results of our numerical simulation. Here, we utilize the general relativistic neutrino radiation hydrodynamics ($\nu$RHD) code, which solves the Boltzmann equations for neutrinos together with the Lagrangian hydrodynamics under spherical symmetry \citep[][]{yamada97, yamada99, sumi05}. We can compute the neutrino transport as well as the dynamics of collapse. We consider four species of neutrino, $\nu_e$, $\bar \nu_e$, $\nu_\mu$ and $\bar \nu_\mu$, assuming that the distribution function of $\nu_\tau$ ($\bar \nu_\tau$) is equal to that of $\nu_\mu$ ($\bar \nu_\mu$). Neutrino oscillation is not taken into account, because it does not occur in the core region and also not affect the dynamics in this stage \citep[][]{chakr11, dasgu12}. A detailed description of the numerical simulations such as general relativistic hydrodynamics, transport and reaction rates of neutrinos can be found in \citet{sumi05}. Referring the resolution dependences shown in \citet{self07}, we use 255 mesh points for the radial Lagrangian coordinate and 20 and 4 mesh points for the energy spectrum and angular distribution, respectively. As the initial conditions, the eight progenitor models described in the previous section are adopted and the outer boundary is settled far from the core region. We utilize an equation of state by \citet{shen98a,shen98b}, which is based on a relativistic mean field model \citep[][]{suga94}, and inhomogeneous matter distribution is described in the Thomas-Fermi approximation \citep[][]{oyak93}.

\begin{figure*}
\plotone{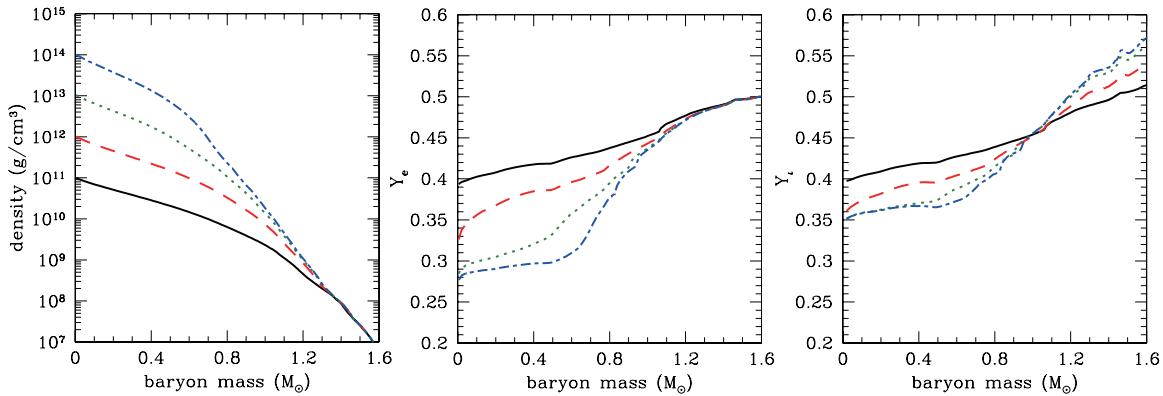}
\caption{Snapshots of the collapsing core for the model with the initial mass $M_\mathrm{init}=13M_\odot$ and the metallicity $Z=0.02$. The left, center and right panels show the density, electron fraction and lepton fraction profiles, respectively. In all panels, solid, dashed, dotted and dot-dashed lines correspond to the times with the central density of $10^{11}$~g~cm$^{-3}$, $10^{12}$~g~cm$^{-3}$, $10^{13}$~g~cm$^{-3}$ and $10^{14}$~g~cm$^{-3}$, respectively.}
\label{prebnc}
\end{figure*}

The density profiles at the time with the central density of $10^{11}$~g~cm$^{-3}$ are shown for all models in Figure~\ref{inidns}. We can recognize that their profiles are very similar for the inner most $1M_\odot$ and the progenitor dependence is seen in the envelope and outer region of the core. The difference will affect the emitted neutrino signal of the collapse-driven supernova. As the density increases due to the collapse, the mean free path of neutrinos gets shorter. In the collapse stage, the emitted neutrinos are mostly $\nu_e$ by the electron capture (\ref{ecp}). Their main opacity source is coherent scattering off nuclei;
\begin{equation}
\nu_e + A \longrightarrow \nu_e + A.
\label{csct}
\end{equation}
When the density exceeds $10^{11 \textrm{-} 12}$~g~cm$^{-3}$, neutrinos are trapped. The neutrino optical depth $d(r)$ at the radius $r$ is defined as
\begin{equation}
d(r) = \int^{R_s}_{r} \frac{dr^{\prime}}{l_\mathrm{mfp}(r^{\prime})},
\label{depth}
\end{equation}
where $R_s$ is the stellar radius and $l_\mathrm{mfp}$ is the mean free path of neutrino. Roughly speaking, the neutrino trapping occurs inside the neutrino sphere, whose radius, $R_\nu$, is defined
as 
\begin{equation}
d(R_\nu) = \frac{2}{3}.
\label{sphere}
\end{equation}
Once the neutrinos are trapped, the inverse process of the electron capture (\ref{ecp}) takes place and $\beta$-equilibrium for the weak interaction is achieved. As a result, the lepton fraction is kept nearly constant and the neutronization is moderated. These trends are shown in Figure~\ref{prebnc} where the profiles of density, electron fraction and lepton fraction are plotted for the model with $(M_\mathrm{init}, Z)=(13M_\odot, 0.02)$. Here, the lepton fraction $Y_l$ is related with the electron fraction $Y_e$ and the electron-type neutrino fraction $Y_{\nu_e}$ as $Y_l=Y_e+Y_{\nu_e}$. Note that, the electron fraction of the inner region decreases even after the neutrino trapping because it is determined by the $\beta$-equilibrium for the given lepton fraction.

The collapse of the core does not halt until the central density exceeds the nuclear density ($\sim \! 3\times10^{14}$~g~cm$^{-3}$). The infalling core is divided into two parts, to which we refer as the inner and outer cores. The inner core contracts subsonically and homologously ($v\propto r$) while the outer core infalls supersonically like free-fall ($v\propto r^{-1/2}$). Roughly speaking, it is known that the inner core mass corresponds to the Chandrasekhar mass $M_\mathrm{Ch} = 0.714(Y_l/0.35)^2 M_\odot$, and does not depend on the progenitors. When the inner core density reaches the nuclear density, nuclei are closely packed and regarded as uniform matter. Then, the adiabatic index increases suddenly due to the repulsive nuclear force and the core restores stability. The influence of nuclear repulsion propagates through the inner core and the collapse is decelerated. On the other hand, since the pressure wave transmits at the speed of sound, the outer core region is still falling supersonically. Therefore, on the boundary of inner and outer cores, the shock wave is formed and launched outward. In the meanwhile, the falling outer core matter is swept by the shock wave and accretes onto the bounced inner core. This compact object is called a proto-neutron star.

\begin{figure}[b]
\plotone{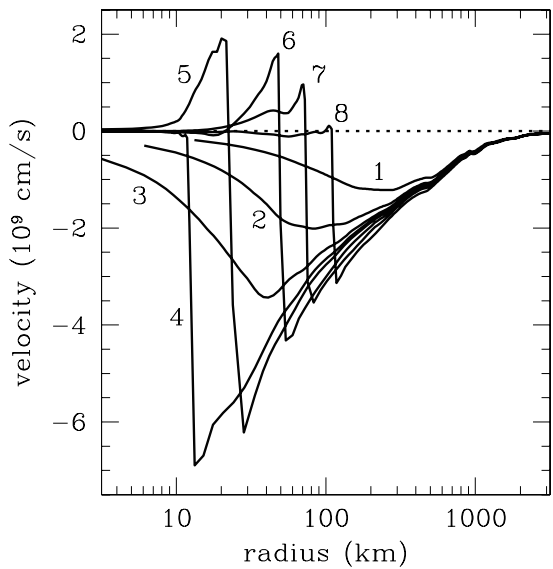}
\caption{Snapshots of velocity profile for the model with the initial mass $M_\mathrm{init}=13M_\odot$ and the metallicity $Z=0.02$. The labels represent the time ordering, where 4 and 8 correspond to the time at the bounce and 5~ms after the bounce, respectively.}
\label{velo}
\end{figure}

A successful explosion is achieved by the shock breakout through the stellar surface. The explosion without the shock stall is called a prompt explosion. However, this scenario is not promising according to recent studies and the shock wave is thought to be stalled. The shock wave propagation is prevented by the photodisintegration reaction of nucleus (\ref{photodis}) and the neutrino emission. As already mentioned, the photodisintegration reaction is an endothermic reaction and consumes the kinetic energy of the shock wave. After the shock wave passes through the neutrino sphere, neutrinos produced by various sorts of reactions can escape carrying out the internal energy behind the shock front. Moreover the shock wave should overcome the ram pressure of the supersonically infalling outer core. The dynamical features described above are seen in Figure~\ref{velo} where the snapshots of velocity profile are shown for the model with $(M_\mathrm{init}, Z)=(13M_\odot, 0.02)$.

The stalled shock wave becomes accretion shock and would revive leading to a supernova explosion in some way. This is called a delayed explosion. Unfortunately, however, the detailed scenario is still an open question. Except the QCD mechanism in which a phase transition to deconfined quark matter makes the second collapse and shock formation and triggers the explosion \citep[][]{fischer11}, spherically symmetric (1D) models are thought to fail. In this case, the ram pressure of accreting matter is higher and prevents the shock revival. On the other hand, if the sphericity is broken, the accretion rate is partially reduced and the explosion has an advantage. In fact, the morphology of collapse-driven supernovae is observationally indicated aspherical \citep[e.g.][]{tanaka09}. Therefore, most of the recently promising scenarios are based on multi-dimensionality.

\begin{figure*}
\plotone{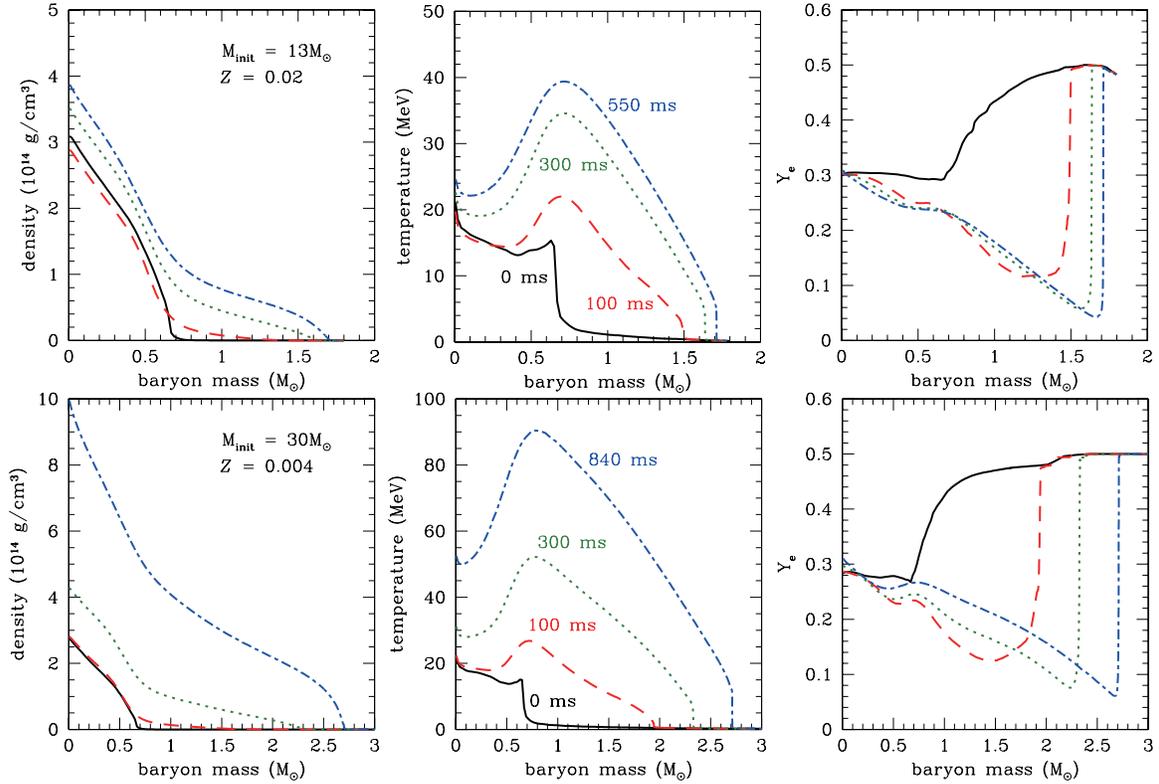}
\caption{Evolutions of the density, temperature and electron fraction profiles by the simulation of neutrino radiation hydrodynamics. The upper plots show the results of the model with the initial mass $M_\mathrm{init}=13M_\odot$ and the metallicity $Z=0.02$, and the lower plots do those of the model with $M_\mathrm{init}=30M_\odot$ and $Z=0.004$. In all panels, solid, dashed and dotted lines correspond to the times at the bounce, 100~ms after the bounce and 300~ms after the bounce, respectively. The dot-dashed lines represent the profiles at 550~ms after the bounce and the profiles at 840~ms after the bounce (2~ms before the black hole formation) for upper and lower panels, respectively.}
\label{pstbnc}
\end{figure*}

The neutrino heating mechanism has been discussed for years. While, as already mentioned, the neutrinos cool the inner region, they contribute the heating just behind the shock front, where free nucleons produced by the photodisintegration reaction (\ref{photodis}) absorb a small part of neutrinos escaping from the inner region. With the aid of this heating, the shock wave is considered to revive on a time scale of the order of 100~ms. This scenario was initially proposed by \citet{bethe85} according to their numerical simulation with 1D model whereas no other group was able to confirm. Recently, employing multi-dimensional simulation, some groups have reported that the onset of the neutrino-driven explosion would be helped by aspherical hydrodynamic turbulence such as the convective instability \citep[e.g.][]{herant94,fryer02,murphy12} and standing accretion shock instability \citep[e.g.][]{blond03,marek09,taki12}. Another candidate for the explosion scenario is the acoustic mechanism \citep[][]{burr06}. The turbulence generated on the accretion shock travels inward and excites oscillation of the proto-neutron star. Then, the acoustic wave originated in the oscillation propagates outward and deposits energy on the accretion shock, which causes the shock revival and explosion. In this hypothesis, the time scale for the shock revival is estimated as $\gtrsim$500~ms, which is longer than that of the neutrino heating mechanism. Magnetic fields may also be important for some supernovae \citep[e.g.][]{lebla70,taki09,ober12}. In fact, neutron stars with strong magnetic fields (magnetars) are observed and their progenitors might also have strong magnetic fields. In the MHD mechanism, the rotational energy converts to the explosion energy via magnetic fields amplified by the field wrapping and/or magnetorotational instability.

While, as discussed above, neutrinos would play a key role for a successful explosion, it is a small part of them that are absorbed near the shock wave and contribute to the explosion. The amount of emitted neutrinos is mainly determined by the released gravitational potential of the accreted matter and thermal energy of the nascent proto-neutron star. Since the neutrino energy deposition for the explosion has an insignificant effect, our $\nu$RHD simulation can be regarded to represent the neutrino signal before the shock revival. In reality, the shock wave would move outward again and the mass accretion stops while when it occurs may depend on the explosion mechanism. However, the shock may not be able to revive if the core is too massive. It is probable for our model with $(M_\mathrm{init}, Z)=(30M_\odot, 0.004)$ and we assume that it fails to explode. In this case, we follow the collapse and neutrino emission up to the black hole formation. For other models, our $\nu$RHD simulations are terminated at 550~ms after the bounce, within which the shock revival is assumed to occur. The evolution after the shock revival is dealt in the next section.

In Figure~\ref{pstbnc}, we show the evolutions of density, temperature and electron fraction after the bounce. They are the results of our $\nu$RHD simulation for the models with $(M_\mathrm{init}, Z)=(13M_\odot, 0.02)$ and $(30M_\odot, 0.004)$. For both models, since the proto-neutron star mass gets larger by the accretion, the density and temperature rise due to gravitational compression. While the density profile is monotonic, the peak of temperature profile resides not at the center but at the medium region. This is because the shock wave does not run from the center and heats the outer matter, as seen in Figure~\ref{velo}. Since the protons created by the photodisintegration reaction (\ref{photodis}) cause the electron capture (\ref{ecp}), the electron fraction decreases for the shocked region. As recognized in the comparison of the models with $(M_\mathrm{init}, Z)=(13M_\odot, 0.02)$ and $(30M_\odot, 0.004)$, the profile at the bounce does not depend on the progenitor because, as already mentioned, the bounced inner core mass does not differ among progenitors. On the other hand, since the accretion rate is determined by the density profile of the outer core, the proto-neutron star mass differs among the models for several times 100~ms after the bounce. As a result, it is also reflected in the proto-neutron star structure. After the shock propagation, the proto-neutron star settles into a hydrostatic configuration. However, the proto-neutron star of the model with $(M_\mathrm{init}, Z)=(30M_\odot, 0.004)$ recollapses suddenly and forms a black hole at 842~ms after the bounce. The qualitative features of black hole formation are same as results in \citet{sumi06,sumi07}

\section{Proto-neutron star cooling} \label{pnsc}

\begin{figure*}
\plotone{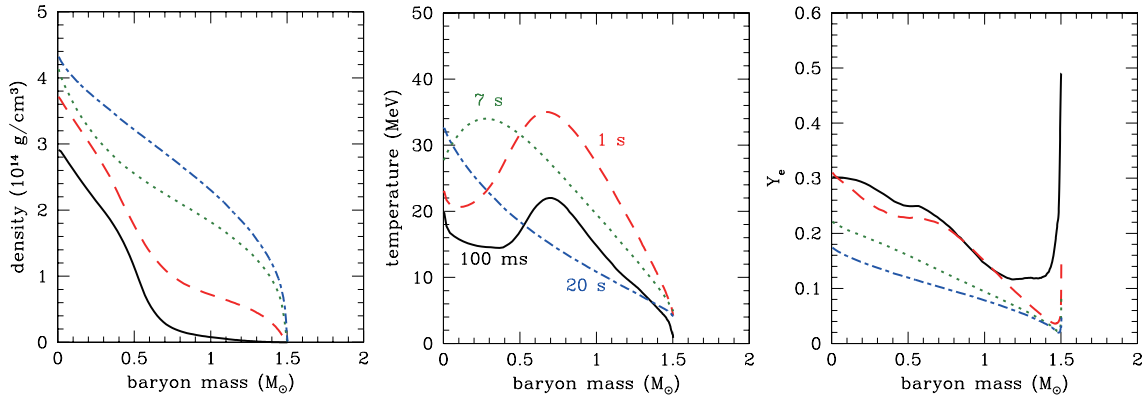}
\caption{Evolutions of the density, temperature and electron fraction profiles by the simulation of proto-neutron star cooling for the model with the initial mass $M_\mathrm{init}=13M_\odot$, the metallicity $Z=0.02$ and the shock revival time $t_\mathrm{revive} = 100$~ms. In all panels, solid, dashed, dotted and dot-dashed lines correspond to the times at 100~ms, 1~s, 7~s and 20~s after the bounce, respectively.}
\label{plpnsc}
\end{figure*}

We evaluate the neutrino luminosities and spectra in the late phase
(after the shock revival) by the proto-neutron star cooling (PNSC)
simulation except for the black-hole-forming case. In this method,
quasi-static evolutions of proto-neutron stars are solved with the
neutrino transfer by multigroup flux limited diffusion scheme under
spherical symmetry with general relativity \citep[][]{suzuki94}. Here, we follow the hydrostatic structure of the proto-neutron star at each time by the Oppenheimer-Volkoff equation while, in $\nu$RHD simulations, the equations for hydrodynamics are fully solved.
We deal the Boltzmann equations in the angle-integrated form for
$\nu_e$, $\bar \nu_e$ and $\nu_x$, where $\nu_\mu$, $\bar \nu_\mu$,
$\nu_\tau$ and $\bar \nu_\tau$ are treated collectively as $\nu_x$. In contrast, $\nu_\mu$ and $\bar\nu_\mu$ are treated individually in our $\nu$RHD simulations. Note that $\nu_\mu$ and $\bar\nu_\mu$ have the same type of reactions and the difference in coupling constants is minor. In fact, the difference between the distribution functions of $\nu_\mu$ and $\bar\nu_\mu$ is typically $\lesssim$1\% in our $\nu$RHD simulations.
The equation of state, binning of neutrino energy and neutrino reactions taken into account are set to the same with the $\nu$RHD simulation shown in the previous section. In this section, we describe the late stage of collapse-driven supernova with our PNSC results. Since the time scale of the neutrino diffusion is $\sim$10~s, the evolution is followed till 20~s after the bounce.

\begin{figure}[b]
\plotone{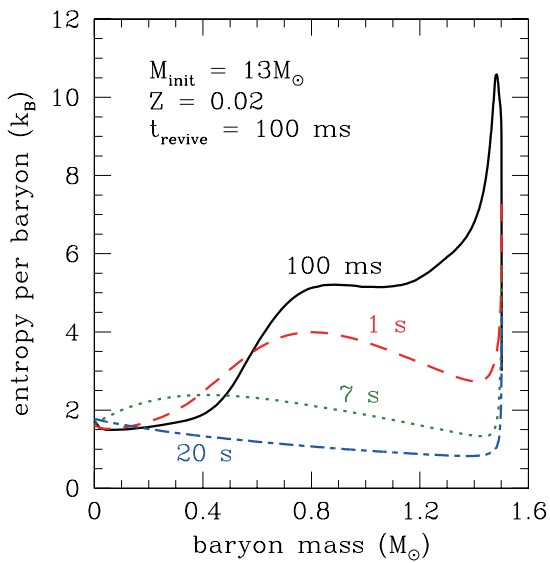}
\caption{Snapshots of entropy profile for the model with the initial mass $M_\mathrm{init}=13M_\odot$, the metallicity $Z=0.02$ and the shock revival time $t_\mathrm{revive} = 100$~ms. The notations of lines is the same as Figure~\ref{plpnsc}.}
\label{etrp}
\end{figure}

We use the results of our $\nu$RHD simulation as initial conditions of
our PNSC simulation. The central parts up to just ahead of the
shock wave are picked up. Here, we take the profiles of electron
fraction and entropy as functions of the baryon mass coordinate from the
$\nu$RHD results and, using them, reconstruct hydrostatic configurations with almost
steady flow of neutrinos, which are used as PNSC initial models. It
is confirmed that, except for the close vicinity of surface, the obtained density profile is consistent with that
of the original $\nu$RHD result as expected because the velocity of
shocked region is negligible (see Figure~\ref{velo}). Since, as already
mentioned, detail of the explosion is not known, we set the shock
revival time $t_\mathrm{revive}$ and $\nu$RHD profiles at
$t_\mathrm{revive}$ are used as PNSC initial conditions. Here, we
investigate three cases as $t_\mathrm{revive}=100$~ms, 200~ms and 300~ms
for each progenitor model. When an explosion mechanism is assumed,
corresponding $t_\mathrm{revive}$ would be determined. Therefore, we can
regard that the explosion scenario described above are parameterized by
the shock revival time. In the PNSC simulation, we follow the evolution
of proto-neutron stars without accretion because amount of matter which falls back onto the proto-neutron star would be minor after the successful explosion.

\begin{figure}[b]
\plotone{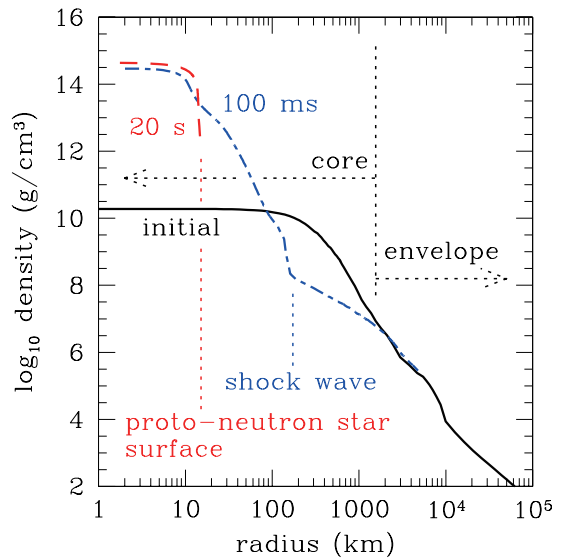}
\caption{Evolution of the density profile for the model with the initial mass $M_\mathrm{init}=13M_\odot$, the metallicity $Z=0.02$ and the shock revival time $t_\mathrm{revive} = 100$~ms. Solid, dot-dashed and dashed line correspond to the onset of collapse, 100~ms after the bounce and 20~s after the bounce, respectively. Dotted vertical lines and horizontal arrows are shown to guide eyes.}
\label{denstot}
\end{figure}

\begin{figure*}
\plotone{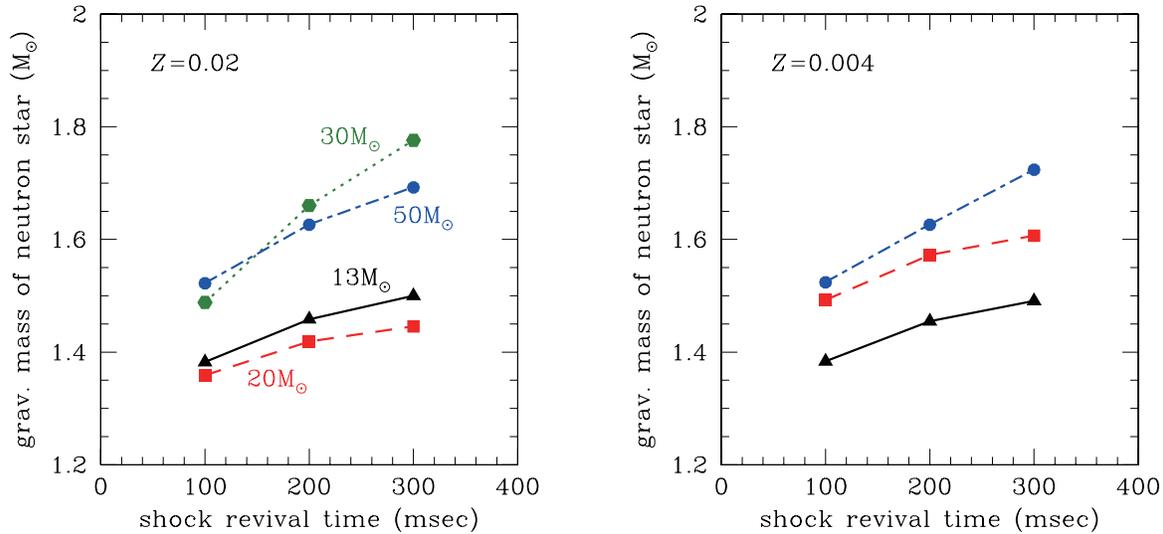}
\caption{Gravitational mass of neutron star models with the metallicity $Z=0.02$ (left panel) and 0.004 (right panel) as a function of the shock revival time. The notations of lines is the same as Figure~\ref{inidns} in both panels.}
\label{nsmass}
\end{figure*}

In Figure~\ref{plpnsc}, we show the evolutions of density, temperature
and electron fraction given by our PNSC simulation for the model with
$(M_\mathrm{init}, Z, t_\mathrm{revive})=(13M_\odot, 0.02,
100~\mathrm{ms})$. One can see that the shocked outer mantle of
proto-neutron star has relatively high temperature and is thermally
expanded at initial moment (solid lines). Neutrinos can easily escape from there
carrying out thermal energy and, therefore, the outer mantle shrinks in a short
time scale to be denser and hotter while the entropy decreases
there (See Figure~\ref{etrp}). Note that diffusion fluxes of $\bar
\nu_e$ and $\nu_x$ initially transport
heat inward from the mantle into the central region in parallel outward to the surface
because number densities of thermal neutrinos have maxima in the hot mantle. This heat flux
attributes to the initial entropy increase in the central region.
Once the temperature profile becomes monotonic, the entire proto-neutron
star cools down gradually. In addition to the neutrino cooling, the net flux
of electron-type neutrinos ($\nu_e - \bar\nu_e$) out of the
proto-neutron star carry away the electron-type lepton number of the
proto-neutron star.
This deleptonization just corresponds to the
neutronization of the proto-neutron star.
The nascent proto-neutron star is lepton-rich and composed also of considerable protons initially. It becomes the ordinary neutron star with less protons achieving the neutrino-less $\beta$ equilibrium in which
there remains small amount of protons and electrons.
The net number flux of electron-type neutrinos is at most $\sim$20\% of the number flux of $\nu_e$.
The decrease of electron fraction shown in
Figure~\ref{plpnsc} represents this process well.
The density evolution from the onset of collapse to 20~s after the core bounce is illustrated in Figure~\ref{denstot}. The features of
proto-neutron star evolution described above are qualitatively common
among the models investigated in this study.

After the revival, the shock wave propagates into the stellar envelope and finally blows it off. Since, in the envelope, the energy losses owing to the photodisintegration and neutrino emission is quite tiny and the binding energy is subtle, the shock wave is not prevented from running outward. While the time scale of the neutrino diffusion is $\sim$10~s, it takes from several hours to days for the shock wave to reach the stellar surface. Since supernova explosion can be observed optically only after the shock breakout, the neutrino signal reaches to the Earth in advance.
Finally, the shock wave blasts through the interstellar medium and forms a supernova remnant, such as the Crab Nebula. On the other hand, a neutron star remains at the center. In Figure~\ref{nsmass}, the gravitational mass of neutron stars $M_{g,\mathrm{NS}}$ considered in our model is plotted. While, according to general relativity, the gravitational mass of a neutron star is different from its baryonic mass $M_{b,\mathrm{NS}}$ (see Table~\ref{keyprm}), there is a one-to-one correspondence between them. We evaluate $M_{g,\mathrm{NS}}$ from the baryonic mass accreted within the shock revival time $t_\mathrm{revive}$ assuming the equation of state by \citet{shen98a,shen98b}. Since the accretion rate is higher, $M_{g,\mathrm{NS}}$ depends on $t_\mathrm{revive}$ especially for the progenitors with larger initial mass $M_\mathrm{init}$. While the neutron star mass could be increased by the accretion from a binary companion, the distribution of neutron star masses may give a hint of the explosion mechanism.

\section{Neutrino signal} \label{neusg}

Neutrinos emitted from the collapse-driven supernova release the gravitational potential of the accreted matter and cool the nascent proto-neutron star. For convenience, we divide the neutrino flux after the shock stall, $F_{\nu_i}(E,t)$, into two terms as
\begin{equation}
F_{\nu_i}(E,t) = F^\mathrm{acc}_{\nu_i}(E,t) + F^\mathrm{cool}_{\nu_i}(E,t),
\label{devide}
\end{equation}
where subscript $i$ denotes the species of neutrinos and $E$ is a neutrino energy. The first term, $F^\mathrm{acc}_{\nu_i}(E,t)$, is an accretion term and related with the accretion luminosity, $L^\mathrm{acc}_\nu(t)$, which is approximated as \citep[][]{thomp03}
\begin{equation}
4\pi R_\mathrm{bnd}^2 \sum_i \int F^\mathrm{acc}_{\nu_i}(E,t) \, dE = L^\mathrm{acc}_\nu(t) \sim \frac{GM_\nu(t) \dot{M}(t)}{R_\nu(t)},
\label{acclumi}
\end{equation}
where $R_\mathrm{bnd}$ is the outer boundary radius of our $\nu$RHD simulation and $R_\nu(t)$ is a radius of the neutrino sphere at the time $t$
defined by equations (\ref{depth}) and (\ref{sphere}). $G$, $\dot{M}(t)$
and $M_\nu(t)$ are the gravitational constant, the mass accretion rate
and the mass enclosed by $R_\nu(t)$, respectively. The second term,
$F^\mathrm{cool}_{\nu_i}(E,t)$, is a cooling term which comes from the
thermal energy loss of the proto-neutron star. While the accretion term is dominant for the early phase ($t \sim 100$~ms), only the cooling term remains after the shock revival leading the explosion.

\begin{figure*}
\plotone{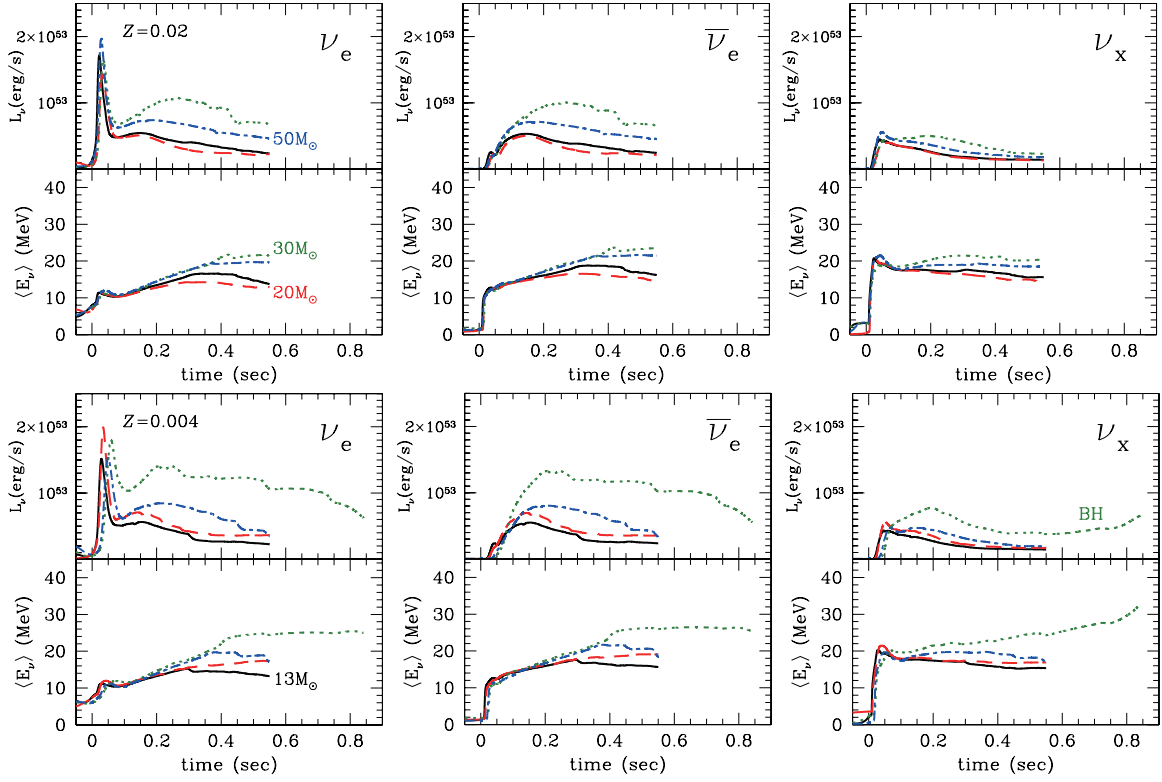}
\caption{Luminosities (upper plots) and average energies (lower plots) of the emitted neutrinos as a function of time after bounce from the $\nu$RHD simulations. The panels correspond, from left to right, to $\nu_e$, $\bar\nu_e$ and $\nu_x$ ($=\nu_\mu$, $\nu_\tau$, $\bar\nu_\mu$, $\bar\nu_\tau$). The results for the models with the metallicity $Z=0.02$ are shown in the top panels, and those for the models with $Z=0.004$ are shown in the bottom panels. In all panels, solid, dashed, dotted and dot-dashed lines correspond to the models with the initial mass $M_\mathrm{init}=13M_\odot$, $20M_\odot$, $30M_\odot$ and $50M_\odot$, respectively. ``BH'' means a black-hole-forming model with $M_\mathrm{init}=30M_\odot$ and $Z=0.004$ and its end point corresponds to the moment of black hole formation.}
\label{nurhd}
\end{figure*}

As already mentioned, supernova explosion is not successful in most of 1D simulations owing to the high mass accretion rate. This would be over estimate because some multi-dimensional effects such as the convective and standing-accretion-shock instabilities reduce the mass accretion in reality. Therefore, we can regard that our 1D $\nu$RHD simulation gives the maximum case of the mass accretion rate. Thus our $\nu$RHD results of the neutrino flux, $F^{\nu\mathrm{RHD}}_{\nu_i}(E,t)$, can be regarded as the upper limit:
\begin{eqnarray}
F^{\nu\mathrm{RHD}}_{\nu_i}(E,t) & = & F^\mathrm{acc,max}_{\nu_i}(E,t) + F^\mathrm{cool}_{\nu_i}(E,t) \nonumber \\
 & \geq & F^\mathrm{acc}_{\nu_i}(E,t) + F^\mathrm{cool}_{\nu_i}(E,t) \nonumber \\
 & = & F_{\nu_i}(E,t),
\label{frhd}
\end{eqnarray}
where $F^\mathrm{acc,max}_{\nu_i}(E,t)$ is the maximum possible value of the accretion term. In Figure~\ref{nurhd}, we show the time profiles of luminosities and average energies of emitted neutrinos evaluated from $\nu$RHD simulation for all models. The end point of the model with initial mass $M_\mathrm{init}=30M_\odot$ and metallicity $Z=0.004$ is a moment of the black hole formation. The peak of $\nu_e$ just after the bounce ($t=0$~s) corresponds to the neutronization burst. When the shock wave propagates through the outer core, nuclei are dissociated into free nucleons and produce a large amount of $\nu_e$ by the electron capture (\ref{ecp}). A short burst of these neutrinos occurs after the shock breakout through the neutrino sphere. This is called a neutronization burst. While the peak luminosity exceeds $10^{53}$~erg~s$^{-1}$, the duration time is the order of 10~ms and the emitted energy is minor comparing with the whole emission of the supernova neutrino.

The persistent emission after the neutronization burst originates from
the mass accretion and proto-neutron star cooling, as described in
equation (\ref{devide}). In this stage, $\nu_e$ and $\bar \nu_e$, which
are abundantly emitted by the electron and positron captures, respectively,
have higher luminosity than $\nu_x$ ($=\nu_\mu=\bar \nu_\mu=\nu_\tau=\bar \nu_\tau$). Nevertheless, the contribution of
$\nu_x$ is not minor because the pair processes such as the
electron-positron pair annihilation, plasmon decay and nucleon
bremsstrahlung occur. Thanks to the shock heating, the accreted matter
is enough hot for electron-positron pair processes. On the other hand,
the average energy of $\nu_x$ is higher than those of $\nu_e$ and $\bar
\nu_e$. Since $\nu_x$ does not have charged-current interactions with
matter consisting of no $\mu^{\pm}, \tau^{\pm}$, its mean free path is longer at the same position than those of $\nu_e$ and $\bar \nu_e$ and its neutrino sphere is smaller as recognized by equations (\ref{depth}) and (\ref{sphere}). Therefore the temperature on the neutrino sphere is higher for $\nu_x$, which makes their average energy also higher. As for the progenitor dependence, since the progenitors with higher density for the range 1.5-$2.0M_\odot$ (see also Figure~\ref{inidns}) have the higher mass accretion rate, their neutrino luminosities are higher as expected from equation (\ref{acclumi}). Their average energies are also somewhat higher but the progenitor dependence is not clear especially for the first $\sim$100~ms. Note that the neutrino emission stops when the proto-neutron star collapses to a black hole for the model with $(M_\mathrm{init}, Z)=(30M_\odot, 0.004)$.

\begin{figure*}
\plotone{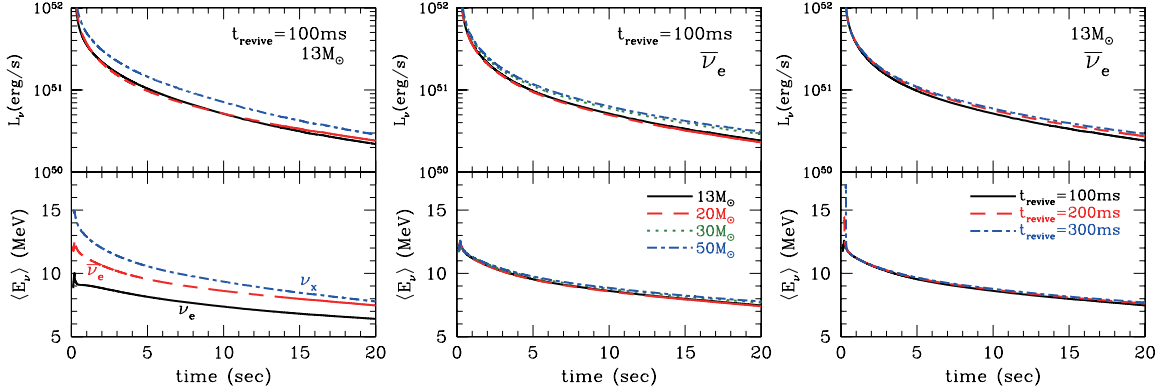}
\caption{Same as Figure~\ref{nurhd} but from the PNSC simulations. In the left panel, signals of $\nu_e$ (solid lines), $\bar\nu_e$ (dashed lines) and $\nu_x$ (dot-dashed lines) are shown for the model with $(M_\mathrm{init}, Z, t_\mathrm{revive})=(13M_\odot, 0.02, 100~\mathrm{ms})$. In the central panel, $\bar\nu_e$ signal are shown for the models with $(Z, t_\mathrm{revive})=(0.02, 100~\mathrm{ms})$ and $M_\mathrm{init}=13M_\odot$ (solid lines), $20M_\odot$ (dashed lines), $30M_\odot$ (dotted lines) and $50M_\odot$ (dot-dashed lines). In the right panel, $\bar\nu_e$ signal are shown for the models with $(M_\mathrm{init}, Z)=(13M_\odot, 0.02)$ and $t_\mathrm{revive} = 100$~ms (solid lines), 200~ms (dashed lines) and 300~ms (dot-dashed lines).}
\label{nupnsc}
\end{figure*}

The results of our PNSC simulation are just corresponding to the cooling
term in equation (\ref{devide}). Obviously, they give the lower limit of
the neutrino flux because accretion induced neutrino flux is not included:
\begin{equation}
F^\mathrm{PNSC}_{\nu_i}(E,t) = F^\mathrm{cool}_{\nu_i}(E,t) \leq F_{\nu_i}(E,t).
\label{fpnsc}
\end{equation}
In Figure~\ref{nupnsc}, we show the time profiles of luminosities and
average energies of emitted neutrinos evaluated from PNSC simulation for
some models.
They decrease in time as the proto-neutron star cools. The neutrino energy hierarchy ($\langle E_{\nu_e} \rangle < \langle E_{\bar \nu_e} \rangle < \langle E_{\nu_x} \rangle$) is same as that in the accretion phase.
The neutrino signals of the models with different initial mass $M_\mathrm{init}$ and shock revival time $t_\mathrm{revive}$ are compared in Figure~\ref{nupnsc}. The luminosity and average energy are higher for the models with larger neutron star mass $M_{g,\mathrm{NS}}$ (see also Figure~\ref{nsmass}) while the difference is not so large.

With inequalities (\ref{frhd}) and (\ref{fpnsc}), we construct the light
curve models of neutrino. For this, we introduce a fraction factor of the accretion term to its maximum, $f(t)$, as a function of time:
\begin{equation}
F^\mathrm{acc}_{\nu_i}(E,t) = f(t) \, F^\mathrm{acc,max}_{\nu_i}(E,t).
\label{ft}
\end{equation}
While $f(t)$ may also depend on the species and energy of neutrino, we ignore their dependences for simplicity. It is required for $f(t)$ to satisfy $f(t) \sim 1$ for the early phase ($t \sim 100$~ms) and $f(t) = 0$ for the phase after the explosion. Using $f(t)$, the neutrino flux is expressed as
\begin{equation}
\begin{array}{rcl}
F_{\nu_i}(E,t) & = & f(t) \, F^\mathrm{acc,max}_{\nu_i}(E,t) + F^\mathrm{cool}_{\nu_i}(E,t) \\
 & = & f(t) \, F^{\nu\mathrm{RHD}}_{\nu_i}(E,t) + (1-f(t)) \, F^\mathrm{PNSC}_{\nu_i}(E,t).
\end{array}
\label{fcon}
\end{equation}
The details of explosion dynamics would determine the function $f(t)$.
For instance, a neutrino signal of the early explosion model corresponds to a rapidly decaying $f(t)$ and a small neutron star mass $M_{g,\mathrm{NS}}$. On the other hand, slowly decaying $f(t)$ and large $M_{g,\mathrm{NS}}$ give a neutrino signal of the late explosion model.

When supernova neutrinos are actually detected, this study would help us to probe the nature of progenitor and remnant. As discussed above, the neutrino luminosity in the accretion phase ($\sim$100~ms after the neutronization burst) is determined by the progenitor model especially for the density profile. The signals in the cooling phase ($\sim$10~s after the neutronization burst) would provide hints for the mass of remnant neutron star. Moreover, if the transition from the accretion phase to the cooling phase is observed, a restriction for the explosion mechanism may be possible. Our results can be hopefully utilized as immediately comparable templates for a neutrino detection.

\begin{figure*}
\plotone{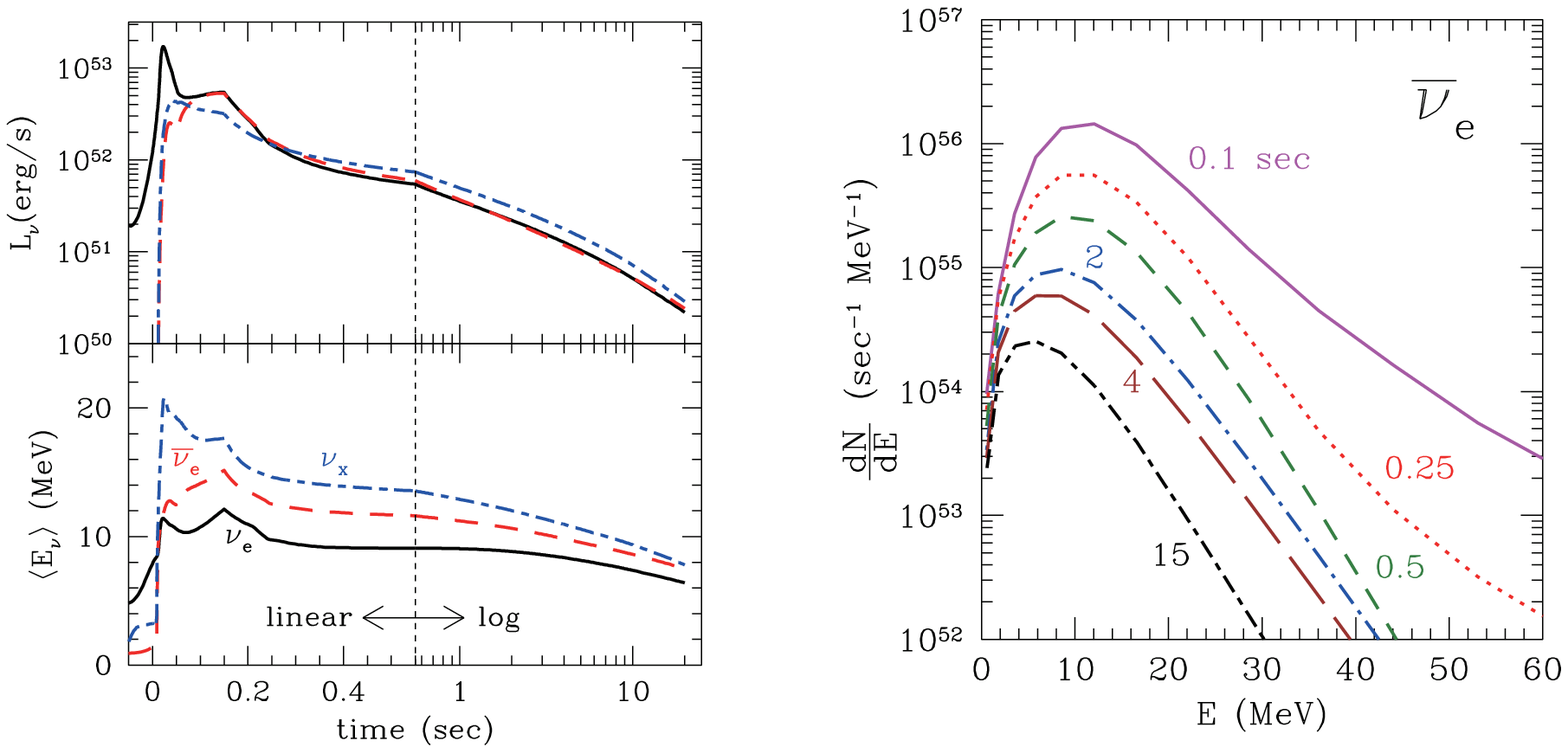}
\caption{Time evolution of neutrino luminosity and average energy (left), and number spectrum of $\bar\nu_e$ (right) from $\nu$RHD and PNSC simulations with the interpolation (\ref{itplft}) for the model with $(M_\mathrm{init}, Z, t_\mathrm{revive})=(13M_\odot, 0.02, 100~\mathrm{ms})$. In the left panel, solid, dashed and dot-dashed lines represent $\nu_e$, $\bar\nu_e$ and $\nu_x$ (dot-dashed lines), respectively. In the right panel, the lines correspond, from top to bottom, to 0.1, 0.25, 0.5, 2, 4 and 15~s after the bounce.}
\label{nucont}
\end{figure*}

\begin{figure}[b]
\plotone{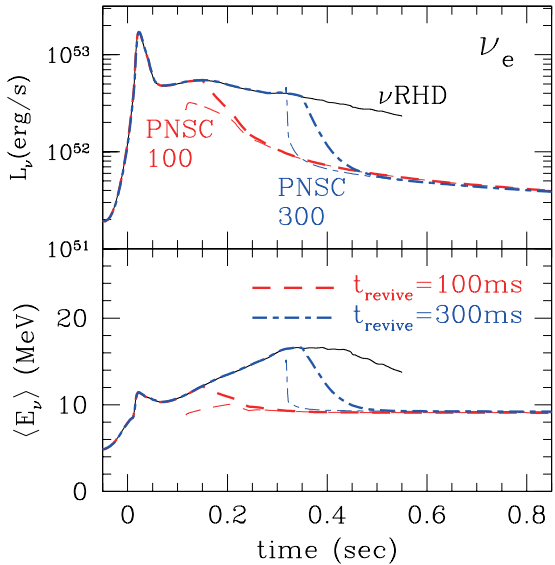}
\caption{Time evolutions of neutrino luminosity and average energy of $\nu_e$ for the model with $(M_\mathrm{init}, Z)=(13M_\odot, 0.02)$. Thick dashed and thick dot-dashed lines represent the interpolations (\ref{itplft}) with $t_\mathrm{revive}=100$~ms and $t_\mathrm{revive}=300$~ms, respectively. Thin solid lines show the results of $\nu$RHD and thin dashed and thin dot-dashed lines do the results of PNSC simulations with $t_\mathrm{revive}=100$~ms and $t_\mathrm{revive}=300$~ms, respectively.}
\label{nujct}
\end{figure}

One may be able to use our results for modeling of the supernova neutrino signals. In Figure~\ref{nucont}, we demonstrate examples of the neutrino light curve and spectrum. They are drawn under the assumption,
\begin{equation}
f(t) = \left\{
\begin{array}{ll}
1, & t \le t_\mathrm{revive}+t_\mathrm{shift}, \\
\exp \left( -\frac{t-(t_\mathrm{revive}+t_\mathrm{shift})}{\tau_\mathrm{decay}} \right), & t_\mathrm{revive}+t_\mathrm{shift} < t,
\end{array}
\right.
\label{itplft}
\end{equation}
for the model with $(M_\mathrm{init}, Z, t_\mathrm{revive})=(13M_\odot, 0.02, 100~\mathrm{ms})$.
The junctions of this interpolation are shown in Figure~\ref{nujct}.
Since $f(t)$ corresponds to the fraction of mass accretion rate to its maximum, the decay time scale $\tau_\mathrm{decay}$ would be a propagation time scale of the revived shock wave.
When the shock revives at the radius $r_\mathrm{shock}$ with the escape velocity $v_\mathrm{esc}(r_\mathrm{shock})$, it takes
\begin{eqnarray}
&& t_\mathrm{propagation}  \sim  \frac{R_\mathrm{core}}{v_\mathrm{esc}(r_\mathrm{shock})} = R_\mathrm{core} \sqrt{\frac{r_\mathrm{shock}}{2GM_{g,\mathrm{NS}}}} \nonumber \\
&& \sim  30~\mathrm{ms} \left(\frac{R_\mathrm{core}}{1000~\mathrm{km}}\right)\left(\frac{M_{g,\mathrm{NS}}}{1.5M_\odot}\right)^{-1/2}\left(\frac{r_\mathrm{shock}}{300~\mathrm{km}}\right)^{1/2}, \nonumber \\
\label{decayt}
\end{eqnarray}
for the shock wave to pass the core with the size of $R_\mathrm{core}$. Here, $r_\mathrm{shock}$ is the most ambiguous parameter but at least $100~\mathrm{km}<r_\mathrm{shock}<1000~\mathrm{km}$ would be satisfied. Therefore, we adopt 300~km as a typical value and get $\tau_\mathrm{decay}=30$~ms. Note that, this value corresponds to also the free-fall time scale of inner core.
The time shift $t_\mathrm{shift}$ is needed for the following two reasons.
First, $F_{\nu_i}(E,t)$ is evaluated as the neutrino flux on the outer boundary of our $\nu$RHD simulation at the time $t$. The outer boundary of our PNSC simulation is the proto-neutron star surface while that of our $\nu$RHD simulation is located in the envelope. The correction of the light traveling time of the distance between two boundaries ($t_\mathrm{travel} \sim O(10)$~ms) is needed.
Second, results of PNSC simulations for very early phase are not reliable because, as already mentioned, the static density profiles are not consistent with those of $\nu$RHD simulations in the close vicinity of surface. It is needed $O(10)$~ms for the relaxation. Therefore we set $t_\mathrm{shift}=50$~ms,
as a typical value. Incidentally, to evaluate the flux numerically, $f(t)$ is settled to zero for $t \ge t_\mathrm{revive}+t_\mathrm{shift}+200$~ms, where $f(t) \ll 1$ in equation (\ref{itplft}).
Hereafter, for simplicity, we take $\tau_\mathrm{decay}=30$~ms and $t_\mathrm{shift}=50$~ms for all cases. Note that, the neutrino number emitted during the period of this interpolation is about 5-15\% of the total amount.

\begin{figure}[b]
\plotone{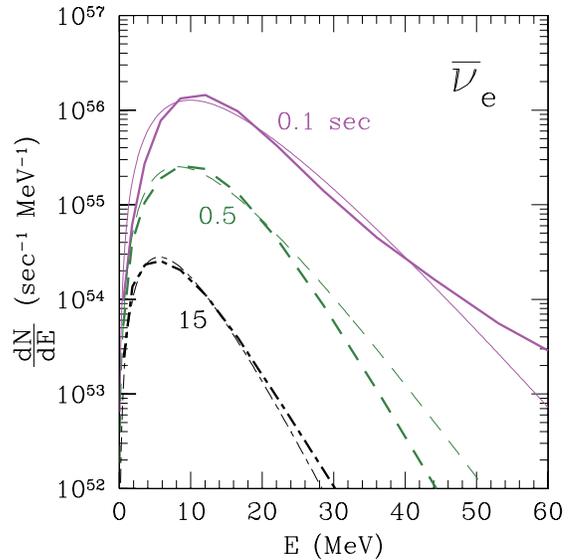}
\caption{Number spectra of $\bar\nu_e$ at selected times for the model with $(M_\mathrm{init}, Z, t_\mathrm{revive})=(13M_\odot, 0.02, 100~\mathrm{ms})$. Solid, dashed and dot-dashed lines correspond to 0.1, 0.5 and 15~s after the bounce, respectively. Thick lines show the results of our simulations while thin lines are Fermi-Dirac spectra with the same luminosity and average energy as the numerical results. The chemical potential is set to zero for the Fermi-Dirac distribution.}
\label{fdfit}
\end{figure}

\begin{figure*}
\plotone{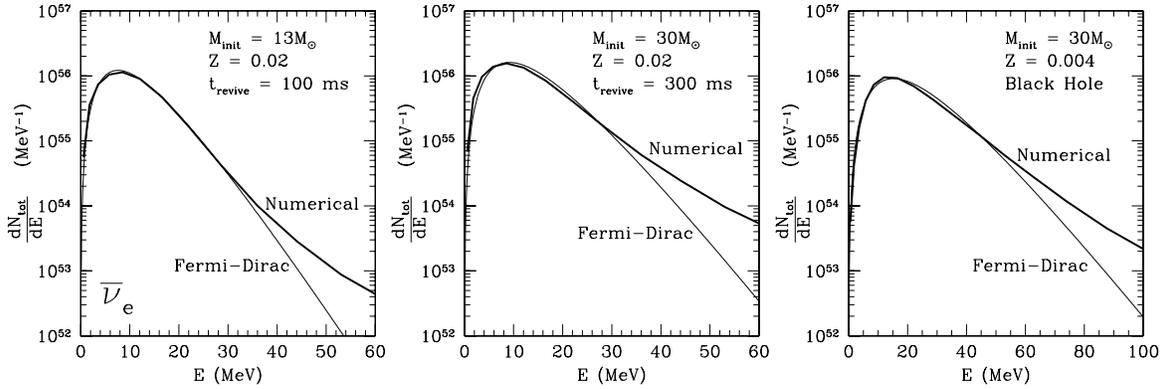}
\caption{Time integrated number spectra of $\bar\nu_e$ over the duration of the simulations. The left, center and right panels correspond to the models of supernova with $(M_\mathrm{init}, Z, t_\mathrm{revive})=(13M_\odot, 0.02, 100~\mathrm{ms})$, supernova with $(M_\mathrm{init}, Z, t_\mathrm{revive})=(30M_\odot, 0.02, 300~\mathrm{ms})$ and black hole formation with $(M_\mathrm{init}, Z)=(30M_\odot, 0.004)$, respectively. The meaning of thick and thin lines are the same as Figure~\ref{fdfit}.}
\label{fdfitint}
\end{figure*}

\begin{figure*}
\plotone{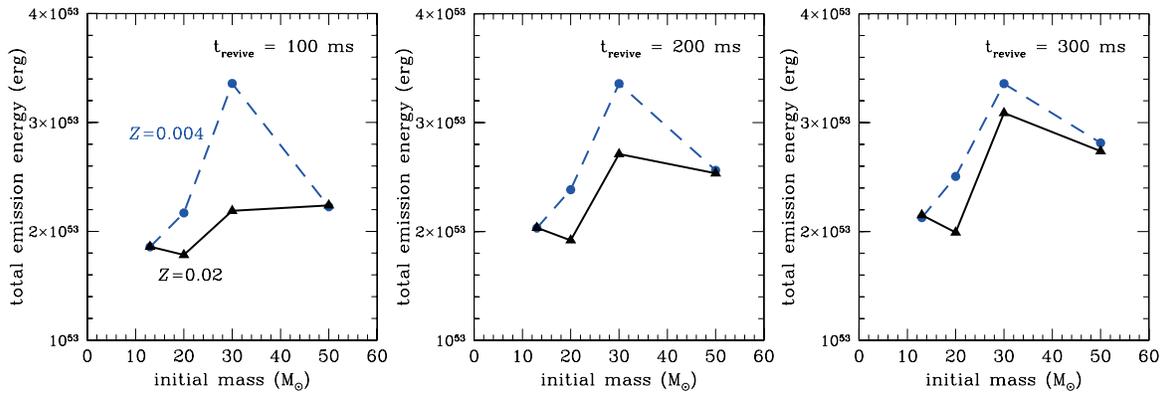}
\caption{Total neutrino energy emitted till 20~s after the bounce for the models with the shock revival time $t_\mathrm{revive} = 100$~ms (left), 200~ms (center) and 300~ms (right). They are computed from $\nu$RHD and PNSC simulations with the interpolation (\ref{itplft}) except for the model with the initial mass $M_\mathrm{init}=30M_\odot$ and the metallicity $Z=0.004$, for which the neutrino emission till the black hole formation followed by $\nu$RHD simulation is plotted in all panels. The notations of lines are the same as Figure~\ref{coremas}.}
\label{etot}
\end{figure*}

In Figure~\ref{fdfit}, the neutrino number spectra are compared with Fermi-Dirac distributions that have the same luminosity and average energy with zero chemical potential.
While Fermi-Dirac spectra roughly fit our results, the deviation is not small.
It is shown that high energy neutrinos are emitted in the accretion phase because the low density outer region has high temperature due to shock heating. On the other hand, after the shock revival, this region becomes dense and the proto-neutron star surface is not heated any more in the cooling phase. Therefore, the deficit of high energy neutrinos can be seen in the spectra of late time.
Figure~\ref{fdfitint} compares, as in Figure~\ref{fdfit}, numerical results and Fermi-Dirac distributions for some cases of time integrated neutrino number spectra over the duration of the simulations. We find that, for the supernova models, the spectrum given by our simulations is relatively well fitted by Fermi-Dirac distribution up to $\sim$30~MeV but has high energy tail originated in the accretion phase. Even for the black-hole-forming model, the Fermi-Dirac distribution roughly fits our spectrum while the numerical uncertainty is larger for the high energy regime.

\begin{figure*}
\plotone{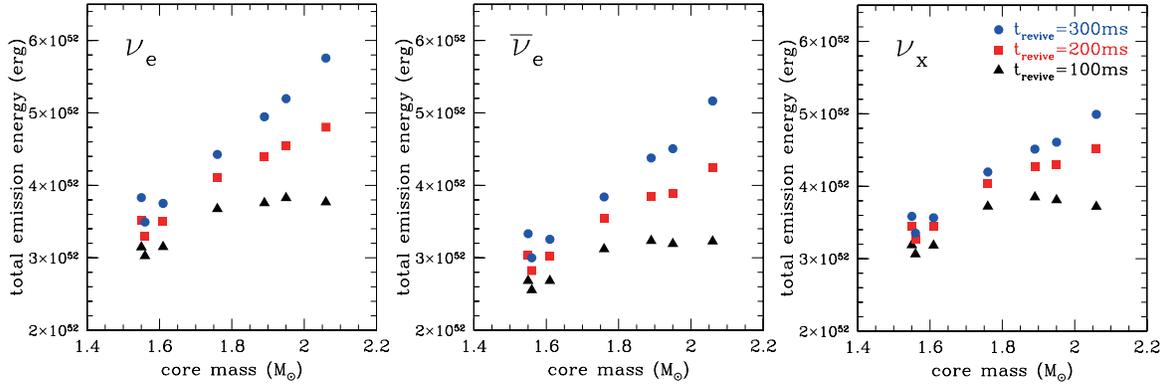}
\caption{Neutrino energy emitted till 20~s after the bounce for $\nu_e$ (left), $\bar\nu_e$ (center) and $\nu_x$ (right). They are computed from $\nu$RHD and PNSC simulations with the interpolation (\ref{itplft}). The plots with triangle, square and circle denote the models with the shock revival time $t_\mathrm{revive} = 100$~ms, 200~ms and 300~ms, respectively.}
\label{etotsp}
\end{figure*}

\begin{figure*}
\plotone{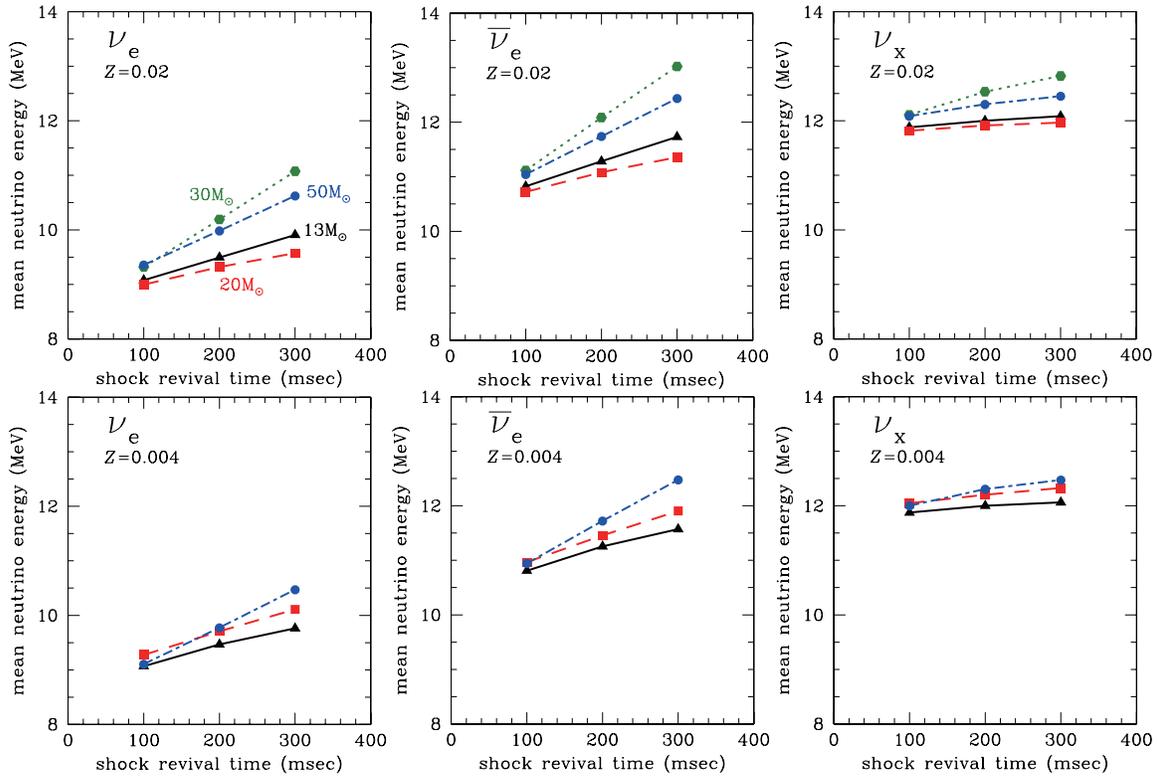}
\caption{Mean energy of emitted neutrinos till 20~s after the bounce for $\nu_e$ (left), $\bar\nu_e$ (center) and $\nu_x$ (right). They are computed from $\nu$RHD and PNSC simulations with the interpolation (\ref{itplft}). The upper and lower panels show the results for models with the metallicity $Z=0.02$ and $Z=0.004$, respectively. The notations of lines are the same as Figure~\ref{nsmass}.}
\label{emean}
\end{figure*}

In Figure~\ref{etot} and Table~\ref{keyprm}, we show the total neutrino energy emitted till 20~s after the bounce for all models with the interpolation (\ref{itplft}). Note that, since the model with $(M_\mathrm{init}, Z)=(30M_\odot, 0.004)$ form a black hole, the total neutrino energy emitted till the black hole formation followed by $\nu$RHD simulation is plotted. We can see that the total emission energy is related to the core mass rather than the initial mass (see also Figure~\ref{coremas}), and it is larger for the explosion models with large $t_\mathrm{revive}$ because the accretion phase is longer and the neutron star mass is larger.
These features are shown more explicitly in Figure~\ref{etotsp}, where the total emission energy is plotted for the core mass. The shock-revival-time dependence of the total emission energy is larger for the models with higher core mass because the mass accretion rate is higher. Since the neutrino luminosity is approximated as equation (\ref{acclumi}), the emission energy during the accretion phase is roughly proportional to the product of the mass accretion rate and shock revival time. Moreover, the resultant proto-neutron star mass depends on this product. Therefore, also in the late phase, the shock-revival-time dependence is larger for the models with higher core mass. On the other hand, the shock-revival-time dependence is larger for $\nu_e$ and $\bar\nu_e$ than $\nu_x$. This feature is also seen for the mean energy of emitted neutrinos, as shown in Figure~\ref{emean}. As already stated, in the accretion phase, $\nu_e$ and $\bar\nu_e$ are emitted more abundantly than $\nu_x$. Therefore the signals of $\nu_e$ and $\bar\nu_e$ is sensitive to the duration of this phase, i.e. the shock revival time.

\section{Summary and Discussion} \label{disc}

The purpose of this study is to construct a comprehensive data set of
long-term (up to $\sim$10~s from the onset of the collapse) supernova
neutrino signal for variety of progenitor stellar models with
different initial masses and metallicities, which would be useful for
a wide range of research related to supernova neutrinos.  To achieve
this goal avoiding the difficulty of long-term full numerical
simulations, we combined two different schemes of numerical
simulations. The early phase of the collapse-driven supernova, at
which the collapsing core is bounced and the shock wave is stalled due
to the matter accretion, has been followed by the general relativistic
neutrino radiation hydrodynamics ($\nu$RHD) code. The late phase after
the shock revival has been dealt by the general relativistic proto-neutron star cooling
(PNSC) simulation which solves quasi-static evolutions with the
neutrino diffusion. The two phases are combined phenomenologically,
taking into account the uncertainty about the explosion mechanism, and
the shock revival time is introduced as a parameter connecting the two
phases. Although this connection is not perfectly consistent as a
single physical simulation, this is currently the best way to follow
supernova neutrino signals up to $\sim$10~s for many progenitor
stellar models. There are still many uncertainties about collapse-driven
supernova physics (e.g., equation of state), but we have chosen the
standard or most popular parameters, to provide theoretical supernova
neutrino emission models expected from the standard picture of
collapse-driven supernovae.  Therefore, the database presented here
would serve as a standard guideline or template for the supernova
neutrino signals.

It is interesting to compare our result with the past calculations of
long-term supernova neutrino emission.  \citet{totani98} presented the
supernova neutrino model till 18~s after the bounce for a progenitor
with $M_\mathrm{init}=20M_\odot$ and investigated the detectability in
detail. In their model, the neutrino luminosity decreases by two
orders of magnitude within $\sim$10~s, which is
similar to our results. However,
their average energy gets higher with time, which contradicts our
results. Recently, \citet{fischer10,fischer12} showed the supernova
neutrino spectra till 20~s after the bounce for a progenitor with
$M_\mathrm{init}=18M_\odot$. While they mimicked explosion with an
artificially enhanced neutrino reaction rate, their results are
qualitatively consistent with ours. In particular, the drops of neutrino
luminosity and average energy due to the onset of shock wave revival,
where the matter accretion vanishes, are also seen in their models. 
Roughly speaking, the light curve by \citet{fischer12} is similar to our models with $t_{\rm revive}=300$~ms.
Incidentally, the neutrino average energy is important for the
nucleosynthesis such as $r$-process and $\nu$-process 
\citep[e.g.][]{woosley90,qian96}.

Because we have calculated many models with different properties of
progenitors in a consistent manner, we can examine the dependence of
supernova neutrino emission properties on progenitors.  It is
quantitatively confirmed that the total emission energy of supernova
neutrinos is related with the core mass of progenitors. Thus it is
larger for the progenitors with lower metallicity, but is not
monotonically related to the initial mass of progenitors due to the
mass loss during the pre-collapse stages. The total neutrino energy
emitted also depends on the shock revival time that determines the
explosion time; it increases with $t_{\rm revive}$ because of more
material accreting to the collapsed core. The increase is $\sim$
20--50\% by changing $t_{\rm revive}$ from 100 to 300~ms.

In the following we discuss some potential applications of the
theoretical supernova neutrino data set for various studies.  First,
our results could be used as immediately comparable templates for the
future detection by neutrino detectors.  As mentioned repeatedly, the
explosion mechanism of collapse-driven supernovae is still
unknown. The detection of neutrinos would give a clue to the diagnosis
of the explosion mechanism, because they come from deep inside the
supernova as shown in this paper. Especially, the drops of luminosity
and average energy of neutrinos are important as the observational
signature of the explosion. Our models would give a quantitative
guideline to search such signatures in future detections. Furthermore,
a sudden stop of neutrino emission would be a signature of a black
hole formation, like our model with $M_\mathrm{init}=30M_\odot$ and
$Z=0.004$, providing an exciting opportunity to directly observe the
birth of a black hole. The neutrino signal reaches to the Earth
earlier than the electromagnetic signal, and hence triggering the
signal by working neutrino detectors and rapidly informing
astronomical communities are of crucial importance.  The various
models presented here would be useful to construct such triggering
systems, especially for optimizing the efficiency around the detection
threshold level.

It would be possible to predict the spectrum of supernova relic
neutrinos using our results because we have evaluated the spectra for
various progenitor models with different initial masses and
metallicities. The supernova relic neutrinos are the integration of
neutrino flux emitted by all collapse-driven supernovae in the
causally-reachable universe. According to the search for supernova
relic neutrinos at Super-Kamiokande \citep[][]{malek03,bays12}, the
signal was still not seen but the upper limit was close to the
standard predictions \citep[][]{hori09}. If the signal of supernova
relic neutrinos is actually detected, it would give us a unique
constraint on the cosmic star formation history and initial mass
function. A significant excess of diffuse supernova neutrino flux
compared with that expected from observed supernova rate may indicate
a contribution from failed supernovae, most likely black hole forming
events, like our model with $M_\mathrm{init}=30M_\odot$ and $Z=0.004$
\citep[][]{luna09,lien10}.

Supernova neutrinos are valuable not only for astrophysics but also
for physics of neutrino itself. While we have not taken into account
the neutrino oscillation in calculations of the models presented
here, it can easily be dealt as a post-process to predict
final neutrino signal reaching detectors on the Earth.
\citep[e.g.][]{kotake06}. Recently, as indicated by the results from
T2K \citep[][]{t2k11} and MINOS experiments \citep[][]{minos11}, the
mixing angle of neutrino oscillation $\theta_{13}$ is confirmed to be
nonzero and evaluated as $\sin^2 2\theta_{13} \sim 0.1$ by the results
from reactor neutrino experiments such as Daya Bay
\citep[][]{dayab12}, RENO \citep[][]{reno12} and Double Chooz
\citep[][]{dchooz12}. Thus, at present, the most undetermined
parameter in the neutrino oscillation is the mass hierarchy. Since,
for $\sin^2 2\theta_{13} \sim 0.1$, the survival probabilities of
$\bar\nu_e$, $\bar\nu_\mu$ and $\bar\nu_\tau$ in the stellar envelope
are different for the normal and inverted mass hierarchies, detections
of the supernova neutrinos would give useful information of the mass
hierarchy \citep[e.g.][]{kotake06}.

The supernova neutrino signal depends on the equation of state,
especially for the black-hole-forming case \citep[][]{sumi06}. The
effect of hyperons, which is not taken into account in our equation of
state \citep[][]{shen98a,shen98b}, would be important not only for the
black hole formation \citep[][]{self12} but also for the proto-neutron
star cooling \citep[][]{keil95}. However, hyperonic equation of state
is an unsettled hot topic. In particular, recently, the mass of the
binary millisecond pulsar J1614-2230 was evaluated as $1.97 \pm
0.04M_\odot$ \citep[][]{demo10}. Unfortunately, this remarkable
precision thanks to a strong Shapiro delay signature excludes almost
all models of hyperonic equation of state, because the maximum mass of
neutron stars gets lower by the hyperon inclusion
\citep[e.g.][]{ishi08,shen11}. Neutrino interactions in matter are
also affected by equation of state. In this study, we have adopted a
single nuclear equation of state of \citet{shen98a,shen98b} without hyperons, which
is based on mostly standard assumptions and can be regarded
as a baseline model. While there are
some issues beyond the scope of this study, we hope that 
the result of this paper will be useful for further progress 
of the related fields in astrophysics and neutrino physics.

\acknowledgments

In this work, numerical computations were partially performed on the supercomputers at Center for Computational Astrophysics (CfCA) in the National Astronomical Observatory of Japan (NAOJ), Research Center for Nuclear Physics (RCNP) in Osaka University, The University of Tokyo, Yukawa Institute for Theoretical Physics (YITP) in Kyoto University, Japan Atomic Energy Agency (JAEA) and High Energy Accelerator Research Organization (KEK). This work was partially supported by Grants-in-Aid for Research Activity Start-up from the Japan Society for Promotion of Science (JSPS) through No.~23840038, and for the Scientific Research on Innovative Areas from the Ministry of Education, Culture, Sports, Science and Technology (MEXT) in Japan through No.~20105004. The authors acknowledge supports by Grants-in-Aid for the Scientific Research from MEXT in Japan through Nos.~22540296 (K.S.) and 24105008 (K.N.).

\appendix
\section{Supernova Neutrino Database} \label{db}
The numerical data of supernova neutrino emission computed in this study are publicly available on the Web at\\
{\tt http://asphwww.ph.noda.tus.ac.jp/snn/}\\
This data set is open for general use in any research for astronomy, astrophysics, and physics. Not only the original data of $\nu$RHD and PNSC simulations but also combined data from the onset of collapse to 20~s after the core bounce with the interpolation (\ref{itplft}) are provided. On the Web, the differential neutrino number flux $\frac{\Delta N_{k,\nu_i}(t_n)}{\Delta E_k}$ and differential neutrino number luminosity $\frac{\Delta L_{k,\nu_i}(t_n)}{\Delta E_k}$ at the time $t_n$ are prepared for each neutrino species $\nu_i$ and energy bin $E_k$. Moreover, the spectral data integrated from the onset of collapse to 20~s after the core bounce with the interpolation (\ref{itplft}) are also shown.

\begin{table*}
\caption{Key Parameters for all Models.} 
\begin{center}
\scalebox{0.88}{
\begin{tabular}{cccccccccccccccc}
 \hline \hline
\multicolumn{1}{c}{} & $M_\mathrm{init}$ & $M_\mathrm{tot}$ & $M_\mathrm{He}$ & $M_\mathrm{CO}$ & $M_\mathrm{core}$ & $t_\mathrm{revive}$ & $M_{b,\mathrm{NS}}$ & $M_{g,\mathrm{NS}}$ & $\langle{E_{\nu_e}\rangle}$ & $\langle{E_{\bar \nu_e}\rangle}$ & $\langle{E_{\nu_x}\rangle}$ & $E_{\nu_e,\mathrm{tot}}$ & $E_{\bar \nu_e,\mathrm{tot}}$ & $E_{\nu_x,\mathrm{tot}}$ & $E_{\nu_\mathrm{all},\mathrm{tot}}$ \\
 $Z$ & ($M_\odot$) & ($M_\odot$) & ($M_\odot$) & ($M_\odot$) & ($M_\odot$) & (ms) & ($M_\odot$) & ($M_\odot$) & (MeV) & (MeV) & (MeV) & ($10^{52}$~erg) & ($10^{52}$~erg) & ($10^{52}$~erg) & ($10^{53}$~erg) \\ \hline
  0.02 & 13 &  12.3 & 3.36 & 1.97 & 1.55 & 100 & 1.50 & 1.39 & 9.08 & 10.8 & 11.9 & 3.15 & 2.68 & 3.19 & 1.86 \\
       &    &       &      &      &      & 200 & 1.59 & 1.46 & 9.49 & 11.3 & 12.0 & 3.51 & 3.04 & 3.45 & 2.03 \\
       &    &       &      &      &      & 300 & 1.64 & 1.50 & 9.91 & 11.7 & 12.1 & 3.83 & 3.33 & 3.59 & 2.15 \\
       & 20 &  17.8 & 5.01 & 3.33 & 1.56 & 100 & 1.47 & 1.36 & 9.00 & 10.7 & 11.8 & 3.03 & 2.56 & 3.06 & 1.78 \\
       &    &       &      &      &      & 200 & 1.54 & 1.42 & 9.32 & 11.1 & 11.9 & 3.30 & 2.82 & 3.27 & 1.92 \\
       &    &       &      &      &      & 300 & 1.57 & 1.45 & 9.57 & 11.4 & 12.0 & 3.49 & 3.00 & 3.35 & 1.99 \\
       & 30 &  23.8 & 8.54 & 7.10 & 2.06 & 100 & 1.62 & 1.49 & 9.32 & 11.1 & 12.1 & 3.77 & 3.23 & 3.72 & 2.19 \\
       &    &       &      &      &      & 200 & 1.83 & 1.66 & 10.2 & 12.1 & 12.5 & 4.80 & 4.24 & 4.51 & 2.71 \\
       &    &       &      &      &      & 300 & 1.98 & 1.78 & 11.1 & 13.0 & 12.8 & 5.76 & 5.16 & 4.99 & 3.09 \\
       & 50 &  11.9 &  --- & 11.9 & 1.89 & 100 & 1.67 & 1.52 & 9.35 & 11.0 & 12.1 & 3.76 & 3.24 & 3.85 & 2.24 \\
       &    &       &      &      &      & 200 & 1.79 & 1.63 & 9.98 & 11.7 & 12.3 & 4.39 & 3.85 & 4.28 & 2.53 \\
       &    &       &      &      &      & 300 & 1.87 & 1.69 & 10.6 & 12.4 & 12.4 & 4.95 & 4.38 & 4.51 & 2.74 \\
 0.004 & 13 &  12.5 & 3.76 & 2.37 & 1.61 & 100 & 1.50 & 1.38 & 9.07 & 10.8 & 11.9 & 3.15 & 2.68 & 3.18 & 1.86 \\
       &    &       &      &      &      & 200 & 1.58 & 1.45 & 9.47 & 11.3 & 12.0 & 3.51 & 3.03 & 3.45 & 2.03 \\
       &    &       &      &      &      & 300 & 1.63 & 1.49 & 9.76 & 11.6 & 12.1 & 3.75 & 3.26 & 3.57 & 2.13 \\
       & 20 &  18.9 & 5.18 & 3.43 & 1.76 & 100 & 1.63 & 1.49 & 9.28 & 11.0 & 12.0 & 3.68 & 3.12 & 3.72 & 2.17 \\
       &    &       &      &      &      & 200 & 1.73 & 1.57 & 9.71 & 11.4 & 12.2 & 4.11 & 3.55 & 4.04 & 2.38 \\
       &    &       &      &      &      & 300 & 1.77 & 1.61 & 10.1 & 11.9 & 12.3 & 4.43 & 3.84 & 4.20 & 2.51 \\
       & 30 &  26.7 & 11.1 & 9.35 & 2.59 & --- &  --- &  --- & 17.5 & 21.7 & 23.4 & 9.49 & 8.10 & 4.00 & 3.36 \\
       & 50 &  16.8 &  --- & 16.8 & 1.95 & 100 & 1.67 & 1.52 & 9.10 & 10.9 & 12.0 & 3.83 & 3.19 & 3.81 & 2.23 \\
       &    &       &      &      &      & 200 & 1.79 & 1.63 & 9.77 & 11.7 & 12.3 & 4.54 & 3.89 & 4.30 & 2.56 \\
       &    &       &      &      &      & 300 & 1.91 & 1.72 & 10.5 & 12.5 & 12.5 & 5.20 & 4.51 & 4.61 & 2.81 \\ \hline
\end{tabular}
}
\end{center}
\label{keyprm}
\tablecomments{$M_\mathrm{init}$ and $Z$ are the initial mass and metallicity of progenitors, respectively. $M_\mathrm{tot}$, $M_\mathrm{He}$ and $M_\mathrm{CO}$ are a total progenitor mass, He core mass and CO core mass when the collapse begins, respectively. Since models with $M_\mathrm{init}=50M_\odot$ become Wolf-Rayet stars, $M_\mathrm{He}$ is not defined and $M_\mathrm{CO}$ equals $M_\mathrm{tot}$. $M_\mathrm{core}$ is a core mass which is defined as the region of oxygen depletion. $t_\mathrm{revive}$ is a shock revival time. $M_{b,\mathrm{NS}}$ and $M_{g,\mathrm{NS}}$ are a baryonic mass and gravitational mass of the remnant neutron stats, respectively. The mean energy of emitted $\nu_i$ till 20~s after the bounce is denoted as $\langle{E_{\nu_i}\rangle} \equiv E_{\nu_i,\mathrm{tot}}/N_{\nu_i,\mathrm{tot}}$, where $E_{\nu_i,\mathrm{tot}}$ and $N_{\nu_i,\mathrm{tot}}$ are the total energy and number of neutrinos. $\nu_x$ stands for $\mu$- and $\tau$-neutrinos and their anti-particles: $E_{\nu_x}=E_{\nu_\mu}=E_{\bar{\nu}_\mu}=E_{\nu_\tau}=E_{\bar{\nu}_\tau}$. $E_{\nu_\mathrm{all},\mathrm{tot}}$ is a total neutrino energy summed over all species. The model with $M_\mathrm{init}=30M_\odot$ and $Z=0.004$ is a black-hole-forming model, for which mean and total neutrino energies emitted till the black hole formation are shown.} 
\end{table*}

\end{document}